\documentclass[floats,floatfix,amssymb,prd,twocolumn,superscriptaddress,nofootinbib,preprintnumbers]{revtex4-2}
\usepackage{graphicx} % Required for inserting images
\usepackage{amssymb,amsmath,amsthm,amsfonts, mathtools,epsfig}
\usepackage[usenames]{color}
\usepackage{xcolor}
\usepackage{epstopdf}

\usepackage{aas_macros}
\usepackage{bm}
\usepackage{dcolumn}
\usepackage{lipsum}
\usepackage{latexsym}
\usepackage{rotating}

\usepackage{enumerate}
\usepackage{tensor,multirow}
\usepackage{url}
\usepackage[linktocpage]{hyperref}

\newcommand{\LL}{\mathcal{L}}

\newcommand{\spatial}[1]{\prescript{(3)}{}{#1}}
\newcommand{\tder}{\partial_{t}}

\newcommand{\UU}{\mathcal U}
\newcommand{\spmet}[1]{\tensor{\gamma}{#1}}
\newcommand{\sproj}[1]{\tensor{h}{#1}}
\newcommand{\nel}{n_{\scriptscriptstyle{EL}}}

\DeclareMathOperator{\diag}{diag}

\newcommand{\sapienza}{Dipartimento di Fisica, Sapienza Università 
	di Roma, Piazzale Aldo Moro 5, 00185, Roma, Italy}
\newcommand{\infn}{INFN, Sezione di Roma, Piazzale Aldo Moro 2, 00185, Roma, Italy}

\begin{document}

\title{Nonlinear photon-plasma interaction and the black hole superradiant instability}

\author{Enrico Cannizzaro}
\email{enrico.cannizzaro@uniroma1.it}
\affiliation{\sapienza}
\affiliation{\infn}

\author{Fabrizio Corelli}
\email{fabrizio.corelli@uniroma1.it}
\affiliation{\sapienza}
\affiliation{\infn}

\author{Paolo Pani}
\email{paolo.pani@uniroma1.it}
\affiliation{\sapienza}
\affiliation{\infn}

% \date{\today}

\begin{abstract}
Electromagnetic field confinement due to plasma near accreting black holes can trigger superradiant instabilities at the linear level, limiting the spin of black holes and providing  novel astrophysical sources of electromagnetic bursts.
However, nonlinear effects might jeopardize the efficiency of the confinement, rending superradiance ineffective.
Motivated by understanding nonlinear interactions in this scenario, here we study the full $3+1$ nonlinear dynamics of Maxwell equations in the presence of plasma by focusing on regimes that are seldom explored in standard plasma-physics applications,  namely a generic electromagnetic wave of very large amplitude but small frequency propagating in an inhomogeneous, overdense plasma. 
We show that the plasma transparency effect predicted in certain specific scenarios is not the only possible outcome in the nonlinear regime: plasma blow-out due to nonlinear momentum transfer is generically present and allows for significant energy leakage of electromagnetic fields above a certain threshold.
We argue that such effect is sufficient to dramatically quench the plasma-driven superradiant instability around black holes even in the most optimistic scenarios.
\end{abstract}

\maketitle

%%%%%%%%%%%%%%%%%%%%%%%%%%%%%%%%%%%%%%%%%%%%%%%%%%%%%%
\section{Introduction}
\label{sec:introduction}
%%%%%%%%%%%%%%%%%%%%%%%%%%%%%%%%%%%%%%%%%%%%%%%%%%%%%%
A remarkable property of black holes~(BHs) is that they can amplify low-frequency radiation in a process known as superradiant scattering~\cite{1971JETPL..14..180Z,Teukolsky:1974yv}, which is the wave analog of energy and angular momentum extraction from a BH through the Penrose's process~\cite{Penrose:1969pc} (see~\cite{Brito:2015oca} for an overview).
In 1972, Press and Teukolsky argued that superradiance can in principle be used to produce a ``BH bomb''~\cite{Press:1972zz} provided the amplified waves are confined in the vicinity of the BH,  leading to repeated scattering and coherent energy extraction.
A natural confining mechanism is provided by the Yukawa decay of massive particles, which is the reason why spinning BHs are unstable against massive bosonic perturbations~\cite{Damour:1976kh,Detweiler:1980uk}.
Due to this superradiant instability~\cite{Brito:2015oca}, massive bosons can extract a significant amount of energy from astrophysical BHs, forming a macroscopic condensate wherein their occupation number grows exponentially.  This phenomenon leads to striking observable signatures, such as gaps in the BH mass-spin distribution and nearly-monochromatic gravitational-wave emission from the condensate~\cite{Arvanitaki:2009fg, Arvanitaki:2010sy}. 
In order for the instability to be efficient, the Compton wavelength of these bosons should be comparable to the BH size, which selects masses around $10^{-11}  (10M_\odot/M)\,{\rm eV}$, where $M$ is the BH mass~\cite{Brito:2015oca}.
The possibility of turning BHs into particle-physics laboratories~\cite{Brito:2014wla} for searches of ultralight dark matter has motivated intense study of the superradiant instability for ultralight spin-0~\cite{Damour:1976kh,Detweiler:1980uk,Zouros:1979iw,Dolan:2007mj,Arvanitaki:2014wva,Arvanitaki:2016qwi,
Brito:2017wnc,Brito:2017zvb}, spin-1~\cite{Pani:2012vp,Pani:2012bp,Witek:2012tr,Endlich:2016jgc,East:2017mrj,East:2017ovw,Baryakhtar:2017ngi, 
East:2018glu,Frolov:2018ezx,Dolan:2018dqv,Siemonsen:2019ebd}, and more recently spin-2~\cite{Brito:2013wya,Brito:2020lup,Dias:2023ynv} fields, which in turn spread into many diverse directions~\cite{Brito:2015oca}.

Already at the very birth of BH superradiance, Press and Teukolsky suggested that in the presence of astrophysical plasma even ordinary photons could undergo a superradiant instability, without the need to invoke beyond Standard Model physics~\cite{Press:1972zz,PressRing}. Indeed, a photon propagating in a plasma acquires an effective mass known as the plasma frequency~\cite{Sitenko:1967,Kulsrud:1991jt,Pani:2013hpa,Conlon:2017hhi} 
\begin{equation}
  \label{eq:plasmafreq}
    \omega_p=\sqrt{\frac{n_{e} e^2}{m}}\approx 10^{-11}\sqrt{\frac{n_e}{10^{-1}\text{cm}^{-3}}}\,\text{eV}/\hbar
\end{equation}
where $n, e$, and $m$ are the plasma density, electron charge, and electron mass, respectively. In the case of interstellar plasma ($n_e \sim 10^{-1}\,{\rm cm}^{-3}$), the effective mass is in the right range to trigger an instability around stellar mass BHs (which was also advocated as a possible explanation for the origin of fast radio
bursts~\cite{Conlon:2017hhi}), whereas primordial plasma could trigger superradiant instabilities in primordial BHs potentially affecting the cosmic microwave background~\cite{Pani:2013hpa}.
Furthermore, astrophysical BHs can be surrounded by accretion disks due to the outward transfer of angular momentum of accreting matter, effectively introducing a geometrically complex plasma frequency.
Given the ubiquity of plasma in astrophysics and the central role that BHs play as high-energy sources and in the galaxy evolution, it is of utmost importance to understand whether the plasma can play an important role in triggering BH superradiant instabilities.

While the first quantitative studies about the plasma-driven superradiant instability~\cite{Pani:2013hpa,Conlon:2017hhi} approximated the dynamics as that of a massive photon with effective mass~\eqref{eq:plasmafreq}, the actual situation is much more complex, since Maxwell's equation must be considered together with the momentum and continuity equation for the plasma fluid.
Recently, a \emph{linearized} version of the plasma-photon system (neglecting plasma backreaction) in curved spacetime was studied in~\cite{Cannizzaro:2020uap, Cannizzaro:2021zbp}, where it was shown that the photon field can be naturally confined by plasma in the vicinity of the BH via the effective mass, forming quasibound states that turn unstable if the BH spins. Nevertheless, a crucial issue was unveiled in~\cite{Cardoso:2020nst}, where it was argued that, during the superradiant phase, \emph{nonlinear} modifications to the plasma frequency turn an initially opaque plasma into transparent, hence quenching the confining mechanism and the instability itself. In the nonlinear regime, a transverse, circularly polarized  electromagnetic~(EM) wave with frequency $\omega$ and amplitude $E$ modifies the plasma frequency of a homogeneous plasma as~\cite{1970PhFl...13..472K} 
%%%
\begin{equation}
    \omega_p=\sqrt{\frac{n_e e^2}{m \sqrt{1+\frac{e^2 E^2}{m^2 \omega^2}}}} \,,\label{omegapnonlin}
\end{equation}
%%%%
where the extra term is the Lorentz factor of the electrons. In other words, as the field grows, the electrons turn relativistic and their relativistic mass growth quenches the plasma frequency. As argued in Ref.~\cite{Cardoso:2020nst}, the threshold of this modification lies in the very early stages of the exponential growth, before the field can extract a significant amount of energy from the BH. 

While in this specific configuration the quenching of the instability is evident, this argument suffers for a number of limitations. In particular, circularly polarized plane waves in a homogeneous plasma are the only solutions that are purely transverse, as the nonlinear $\vec{v} \times \vec{B}$ Lorentz force vanishes (here $\vec{v}$ is the velocity of the electron, while $\vec{B}$ is the magnetic field). In this case, the plasma density is not modified by the travelling wave and even a low-frequency wave with large amplitude can simply propagate in the plasma, without inducing a nonlinear backreaction. 
In every other configuration instead (including an inhomogeneous plasma, different polarization, or breaking of the planar symmetry, all expected for setups around BHs), longitudinal and transverse modes are coupled, and therefore the plasma density can be dramatically modified by the  propagating field. This backreaction effect leads to a richer phenomenology as high-amplitude waves can push away electrons from some regions of the plasma, thus creating both a strong pile-up of the electron density in some regions and a plasma depletion in other regions. For example, in the case of a circularly polarized wave scattered off an inhomogeneous plasma, the backreaction on the density increases the threshold for relativistic transparency, as electrons are piled up in a narrow region, thus increasing the local density and making nonlinear transparency harder~\cite{Cattani}. However, in the case of a coherent long-timescale phenomenon such as superradiant instability, one might expect that, if the plasma is significantly pushed away by a strong EM field, the instability is quenched a priori, regardless of the transparency. Overall, the idealized configuration of Ref.~\cite{1970PhFl...13..472K} never applies in the superradiant system, and the nonlinear plasma-photon interaction is much more involved.

The goal of this work is to introduce a more complete description of the relevant plasma physics needed to understand plasma-photon interactions in superradiant instabilities. To this purpose, we shall perform $3+1$ nonlinear numerical simulations of the full Maxwell's equations. 
Clearly, this is a classical topic in plasma physics~\cite{RevModPhys.81.1229,PhysRevLett.64.2011}. Here we are interested in a regime that is relevant for BH superradiance but is seldom studied in standard plasma-physics applications, namely a low-frequency, high-amplitude EM wave propagating in an inhomogeneous overdense plasma.

\section{Field Equations}

For simplicity, and because the stress-energy tensor of the plasma and EM field is negligible even during the superradiant growth, we shall considered a fixed background and neglect the gravitational field. We consider a system composed by the EM field and a plasma fluid, described by the field equations (in rationalized Heaviside units with $c = 1$):
\begin{align}
	\nabla_{\mu} F^{\mu\nu} &= J^\nu, 			\label{eq:FieldEM} \\
	u^\nu \nabla_\nu u^\mu &= \frac{e}{m} F^{\mu\nu} u_\nu,	\label{eq:FieldPlasmaEvolution} \\
	\nabla_\mu(n_e u^\mu) &= 0,				\label{eq:FieldPlasmaContinuity}
\end{align}
where $F_{\mu\nu}$ is the EM tensor, $J^\mu$ is the EM 4-current, $u^\mu$ is the 4-velocity field for the plasma fluid, and $n_e$ is the rest number density of electrons inside the plasma.

Having in mind future extensions, below we perform
a $3+1$ decomposition\footnote{We shall use Greek alphabet to denote spacetime indices $\mu, \nu \in \{0, 1, 2, 3\}$, and Latin alphabet to denote spatial indices $i, j \in \{1, 2, 3\}$.} of the field equations that is valid for any curved background spacetime. However, in this work we will perform our simulations in flat spacetime, $ds^2 = \eta_{\mu\nu} \, dx^\mu \, dx^\nu$.

\subsection{$3+1$ decomposition of the field equations} \label{sec:3+1}

\subsubsection{Generic spacetime} \label{sec:3+1:GenericSpacetime}
Let us introduce a foliation of the spacetime into spacelike hypersurfaces $\Sigma_t$, orthogonal to the 4-velocity of the Eulerian observer $n^\mu$. We then express the line element as
\begin{equation}
    ds^2 = -(\alpha^2 - \beta_i \beta^i) \, dt^2 + 2 \, \beta_i \, dx^i \, dt + \spmet{_{ij}} \, dx^i \, dx^j,
    \label{eq:LineElement}
\end{equation}
where $\alpha$ is the lapse, $\beta^i$ is the shift vector, and $\spmet{_{ij}}$ is the spatial 3-metric.
We can define the electric and the magnetic fields as~\cite{Alcubierre:2009ij}
\begin{align}
	E^\mu = -n_\nu F^{\nu\mu},		\qquad
	B^\mu = -n_\nu F^{\ast \nu\mu},	\label{eq:EBDef}
\end{align}
where $F^{\ast \mu\nu}= -\frac{1}{2} \epsilon^{\mu\nu\lambda\sigma} F_{\lambda\sigma}$ is the dual of $F^{\mu\nu}$. The EM tensor can be decomposed as
\begin{equation}
	F^{\mu\nu} = n^\mu E^\nu - n^\nu E^\mu + \spatial{\epsilon}^{\mu\nu\sigma} B_\sigma,
	\label{eq:FmunuDecomposed}
\end{equation}
where $\spatial{\epsilon}^{\mu\nu\sigma} = n_\lambda \epsilon^{\lambda \mu\nu\sigma}$ is the Levi-Civita tensor of the spacelike hypersurface $\Sigma_t$. Note that $E^\mu$ and $B^\mu$ are orthogonal to $n^\mu$ and are spacelike vectors on the 3-surfaces $\Sigma_t$.

We can define the charge density as $\rho = n_\mu J^\mu$, and the 3-current as $\spatial{J}^\mu = \sproj{^\mu_\nu} J^\nu$, where $\sproj{^\mu_\nu}$ is the projection operator onto $\Sigma_t$. Finally, we can write the Maxwell equations as~\cite{Alcubierre:2009ij}
\begin{align}
	D_i E^i &= \rho,											\label{eq:Gauss} \\
	D_i B^i &= 0,												\label{eq:MagneticGauss} \\
	\tder E^i &= \LL_\beta E^i + \alpha K E^i + [\vec{D} \times (\alpha \vec{B})]^i + \alpha \spatial{J}^i,	\label{eq:EiEvol} \\
	\tder B^i &= \LL_\beta B^i + \alpha K B^i - [\vec{D} \times (\alpha \vec{E})]^i,			\label{eq:BiEvol}
\end{align}
where $D_i$ is the covariant derivative with respect to the 3-metric $\gamma_{ij}$, and $K_{ij}$ is the extrinsic curvature.
Here the first equation is the Gauss' law, the second equation is equivalent to the absence of magnetic monopoles, and the last two are the evolution equations for the electric and magnetic fields, respectively. 
The EM 4-current is given by ions and electrons $J^\mu = J^\mu_{\rm(ions)} + J^\mu_{(e)}$. We assume ions to be at rest, due to the fact that $m \ll m_{\rm(ions)}$, so that $J^\mu_{\rm(ions)} = - \rho_{\rm(ions)} n^\mu$. For electrons instead we have $J^\mu_{(e)} = -e n_e u^\mu$. 
Let us decompose $u^\mu$ into a component along $n^\mu$, $\Gamma = -n^\mu u_\mu$, and a component on the spatial hypersurfaces, $\spatial{u}^\mu = \sproj{^\mu_\nu} u^\nu$. The 4-velocity of the fluid can be written as
\begin{equation}
	u^\mu = \Gamma n^\mu + \spatial{u}^\mu = \Gamma (n^\mu + \UU^\mu)\,, \label{eq:uDecomposition}
\end{equation}
where we defined $\spatial{u}^{\mu} = \Gamma \UU^\mu$.
The above expression allows us to write $\rho = n_\mu J^\mu = \rho_{\rm(ions)} + \rho_{(e)} = \rho_{\rm(ions)} + e \nel$, where $\nel = \Gamma n_e$ is the electron density as seen by the Eulerian observer. The density of ions is constant in time, and will be fixed when constructing the initial data\footnote{Note that with the conventions we used, electrons carry \textit{positive} charge, while ions carry \textit{negative} charge.}. 
As $J^\mu_{\rm(ions)}$ is orthogonal to $\Sigma_t$, the 3-current $\spatial{J}^\mu$ receives only contributions from electrons, and we have $\spatial{J}^\mu = - e n_e \Gamma \UU^\mu = -e \nel \UU^\mu$. Thus, the source terms that appear in Eqs.~\eqref{eq:Gauss}-\eqref{eq:EiEvol} are
\begin{align}
	\rho = \rho_{\rm(ions)} + e \nel, \qquad
	\spatial{J}^\mu = - e \nel \UU^\mu. \label{eq:EMCurrent}
\end{align}

Let us now move to Eq.~\eqref{eq:FieldPlasmaEvolution}. Projecting it on $n^\mu$ and $\Sigma_t$ we obtain respectively (see Appendix~\ref{app:3+1} for the explicit computation):
\begin{align}
	\tder \Gamma &= \beta^i \partial_i \Gamma - \alpha \UU^i \partial_i \Gamma + \alpha \Gamma K_{ij} \UU^i \UU^j \nonumber \\
                 &- \Gamma \UU^i \partial_i \alpha + \frac{e}{m} \alpha E^i \UU_i \, ,	\label{eq:GammaEvolGeneric} \\
    \tder \UU^i &= \beta^j \partial_j \UU^i - \UU^j \partial_j \beta^i - \alpha a^i - \alpha \UU^i K_{jl} \UU^j \UU^l \nonumber \\
                &+ \frac{\alpha}{\Gamma} \frac{e}{m} \Bigl( - \UU^i E^j \UU_j + E^i + \tensor{\spatial{\epsilon}}{^{i j l}} B_l \UU_j \Bigr) \nonumber \\
                &+ 2 \alpha \tensor{K}{^i_j} \UU^j + \UU^i \UU^j \partial_j \alpha - \alpha \UU^j D_j \UU^i \label{eq:UUEvolGeneric}
\end{align}
Finally, we can write the continuity equation~\eqref{eq:FieldPlasmaContinuity} as
\begin{equation}
    \tder \nel = \beta^i \partial_i \nel  + \alpha K \nel  - \alpha \UU^i \partial_i \nel - \alpha \nel \nabla_\mu \UU^\mu.
    \label{eq:NELEvolGeneric}
\end{equation}
While the above decomposition is valid for a generic background metric, from now on we will focus on a flat spacetime.

%%%%%%%%%%%%%%%%%%%
\subsubsection{Flat spacetime} \label{sec:3+1:Flat Spacetime}
%%%%%%%%%%%%%%%%%%%
We use Cartesian coordinates, so that $g_{\mu\nu} = \eta_{\mu\nu} = \diag \{-1, 1, 1, 1\}$. As a consequence, we have that for any 3-vector $\spatial{V}^i = \spatial{V}_i$, and
\begin{align}
	\alpha = 1, \qquad
	\beta^i = 0, \qquad
	K_{ij} = 0.
	\label{eq:MinkowskiMetricAlphaBetaK}
\end{align}

In these coordinates we can write the equations for the EM field as    
\begin{align}
	\partial_i E^i &= \rho_{\rm(ions)} + e \nel,					\label{eq:GaussCartesian} \\
	\partial_i B^i &= 0,								\label{eq:MagneticGaussCartesian} \\
	\tder E^i &= [\vec{\partial} \times \vec{B}]^i - e \nel \UU^i,	\label{eq:EiEvolCartesian} \\
	\tder B^i &= - [\vec{\partial} \times \vec{E}]^i,			\label{eq:BiEvolCartesian}
\end{align}
the evolution equations for $\Gamma$ and $\UU^i$ as
\begin{align}
	\tder \Gamma &= - \UU^i \partial_i \Gamma + \frac{e}{m} E^i \UU_i,	\label{eq:GammaEvolCartesian} \\
	\tder \UU^i &= - \UU^j \partial_j \UU^i + \frac{1}{\Gamma} \frac{e}{m} \Bigl[ -\UU^i E^j \UU_j + E^i + (\vec{\UU} \times \vec{B})^i \Bigr],	\label{eq:UUEvolCartesian}
\end{align}
and the continuity equation as
\begin{equation}
	\tder \nel = - \UU^i \partial_i \nel - \nel \partial_i \UU^i.
	\label{eq:NELEvolCartesian}
\end{equation}

Moreover, from the normalization condition that $u^\mu u_\mu = -1$ we can obtain a constraint for $\Gamma$ and $\UU^i$:
\begin{equation}
	\Gamma^2(1 - \UU^i \UU_i) = 1.
	\label{eq:PlasmaConstraintCartesian}
\end{equation}

\section{Numerical Setup}  \label{sec:NumericalSetup}
In this section we discuss our numerical setup, describing the integration scheme and the initialization procedure.

\subsection{Integration scheme} \label{sec:IntegrationScheme}

The system of evolution equations can be schematically written in the form
\begin{equation}
    \tder Y = S[Y, \vec \partial Y],
\label{eq:EvolutionSystem}
\end{equation}
where $Y = (\vec{E}, \vec{B}, \Gamma, \vec \UU, \nel)$ is the vector containing the fields, while $S$ comprises the right hand sides of Eqs.~\eqref{eq:EiEvolCartesian}-\eqref{eq:NELEvolCartesian}, and depends on the fields and their spatial derivatives. Note that  $\rho_{\rm(ions)}$ is not included in $Y$, since it is kept constant during the evolution, consistently with the approximation that ions are at rest.

We perform the numerical integration of the system \eqref{eq:EvolutionSystem} with the method of lines. We use the fourth-order accurate Runge-Kutta method for time evolution, computing the spatial derivatives of the fields in $S$ with the fourth-order accurate centered finite differences scheme. In order to monitor the accuracy and the reliability of our integration algorithm, we evaluate the violation of the constraints~\eqref{eq:GaussCartesian} and~\eqref{eq:PlasmaConstraintCartesian}. Namely, we define the quantities
\begin{align}
CV_{\rm Gauss} &= \partial_i E^i - e \nel - \rho_{\rm(ions)}, \label{eq:CVGauss} \\
CV_{\rm Plasma} &= \sqrt{\Gamma^2(1 - \UU_i \UU^i)} - 1, \label{eq:CVPlasma}
\end{align}
and we check that increasing the resolution they decrease to zero as fourth order terms, consistently with the accuracy of our integration scheme. The results of our convergence tests are shown in Appendix~\ref{app:Convergence}.

% We evolve $\vec{E}$, $\vec{B}$, $\Gamma$, $\vec{\UU}$, and $\nel$ with Eqs.~\eqref{eq:EiEvolCartesian}-\eqref{eq:NELEvolCartesian}, using the constraints~\eqref{eq:GaussCartesian} and~\eqref{eq:PlasmaConstraintCartesian} to evaluate the convergence of the code. The profile of $\rho_{\rm(ions)}$ is kept constant, consistently with the approximation that ions are at rest. 
% For the numerical integration we used the fourth-order accurate Runge-Kutta algorithm, computing the spatial derivatives with the fourth-order accurate centered finite differences scheme. 
For simplicity we shall simulate the propagation of \emph{plane} EM wave packets along the $z$ direction, and therefore we will obtain field configurations that are homogeneous along the $x$ and $y$ directions. This feature allows us to impose periodic boundary conditions in the $x$ and $y$ directions, as they preserve the homogeneity of the solution without introducing numerical instabilities. We impose periodic boundary conditions also on the $z$ axis and, in order to avoid the spurious interference of the EM wave packet with itself, we choose grids with extension along $z$ large enough to avoid interaction with spurious reflected waves during the simulations. 

\subsection{Initialization procedure} \label{sec:Initialization}

\begin{figure}
	\includegraphics[width = \columnwidth]{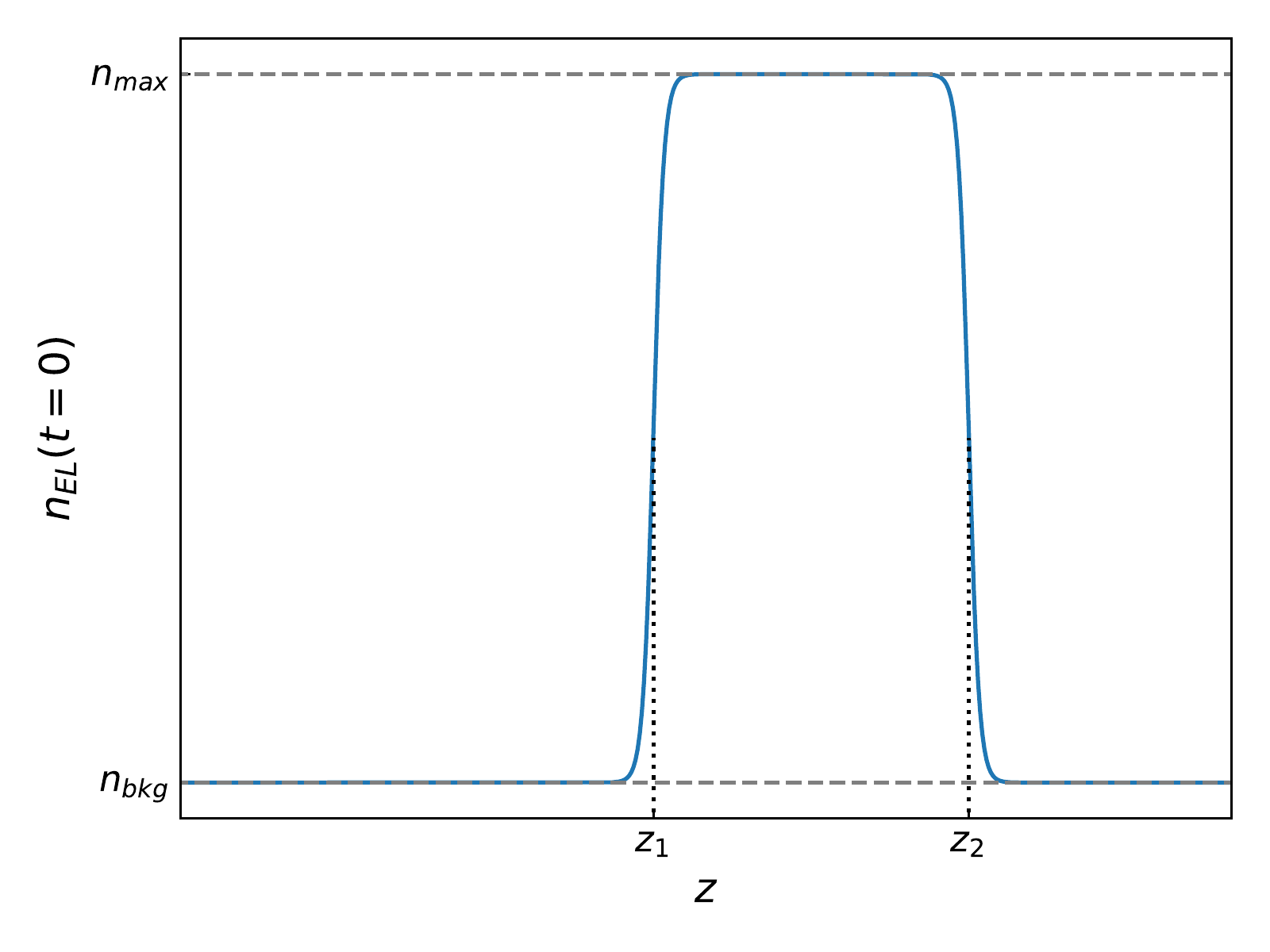}
	\caption{Qualitative behavior of the barrier-shaped initial profile for the plasma, Eq.~\eqref{eq:SigmoidWall}. $n_{\rm max}$ and $n_{\rm bkg}$ are the values of the plasma density inside the barrier and on the background, while the parameters $z_{1, 2}$ and $W_{1,2}$ determine the position and the steepness of the boundaries of the barrier, respectively.}
	\label{fig:SigmoidWall}
\end{figure}

When constructing the initial data for the simulations we first set the profile of the plasma.
We start by setting $\Gamma(t = 0, \vec{x}) = 1$ and $\vec{\UU}(t = 0, \vec{x}) = 0$, so that the plasma is initially at rest. Then, we initialize the profile of $\nel$ with barrier-like shape of the following form:
\begin{align}
	\nel(t = 0, \vec{x}) &=  2 n_{\rm bkg} - n_{\rm max} \nonumber \\
                         &+ (n_{\rm max} - n_{\rm bkg}) \nonumber \\
                         &\times\Bigl[\sigma(z; W_1, z_1) + \sigma(z; -W_2, z_2)\Bigr],
	\label{eq:SigmoidWall}
\end{align}
Where $\sigma(z; W, z_0) = (1+e^{-W(z - z_0)})^{-1}$ is a sigmoid function. The qualitative behavior of Eq.~\eqref{eq:SigmoidWall} is shown in Fig.~\ref{fig:SigmoidWall}, where we can see that $n_{\rm bkg}$ is the background value of the plasma density and $n_{\rm max}$ is the plasma density at the top of the barrier. The parameters $z_{1,2}$ determine the location and width of the barrier, while the parameters $W_{1, 2}$ control its steepness. Note that this profile was chosen to reproduce a very crude toy model of a matter-density profile around a BH~\cite{Novikov:1973kta}, where the accretion flow peaks near the innermost stable circular orbit and is depleted between the latter and the BH horizon. 
In our context this configuration is particularly relevant because EM waves can be superradiantly amplified near the BH and plasma confinement can trigger an instability~\cite{Pani:2013hpa,Cardoso:2013fwa, Dima:2020rzg, Cannizzaro:2020uap,Cannizzaro:2021zbp,Lingetti:2022psy,Wang:2022hra}.
Finally, the constant profile of $\rho_{\rm(ions)}$ is determined by imposing that the plasma is initially neutral, so that $\rho_{\rm(ions)}(t = 0, \vec{x}) = -e \nel(t = 0, \vec{x})$. 
Once the profile of the plasma has been assigned we proceed to initialize the EM field. We consider a circularly polarized wave packet moving forward in the $z$-direction:
\begin{align}
	\vec{E} &= A_E
	\begin{pmatrix}
		\cos[k_z(z - z_0)]\\
		\sin[k_z(z - z_0)]\\
		0
	\end{pmatrix}
	e^{-\frac{(z - z_0)^2}{2 \sigma^2}}, \\
	\vec{B} &= A_E \frac{k_z}{\omega}
	\begin{pmatrix}
		- \sin[k_z(z - z_0)]\\
		\cos[k_z(z - z_0)]\\
		0
	\end{pmatrix}
	e^{-\frac{(z - z_0)^2}{2 \sigma^2}},
	\label{eq:EMInitial}
\end{align}
where $A_E$ is the amplitude of the wave packet, $\sigma$ is its width, $z_0$ its central position, $\omega$ is the frequency, and $k_z = \sqrt{\omega^2 - \omega_p^2}$, where $\omega_p = \sqrt{\frac{e^2 \, n_{\rm bkg}}{m}}$ is the plasma frequency computed using $n_{\rm bkg}$, as the wave packet is initially located outside the barrier (i.e., $\sigma\ll z_1-z_0$).

\section{Results}   \label{sec:results}

Here we present the results of our numerical simulations of nonlinear plasma-photon interactions in different configurations. We shall consider a low-frequency, circularly polarized wave packet propagating along the $z$ direction and scattering off the plasma barrier with the initial density profile given by Eq.~\eqref{eq:SigmoidWall}.

\subsection{Linear regime}    \label{sec:results:linear}

As a consistency check of our code, we tested that for sufficiently low amplitude waves our simulations are in agreement with the predictions of linear theory.
We set units such that $e = m = 1$ and consider an initial wave packet of the electric field centered at $z_0 = 0$, with a characteristic width $\sigma  = 5$. We also set $\omega = 0.5$ and $A_E =  10^{-6}$, so that the evolution can be described by the linear theory. The plasma barrier was situated between $z_1 = 40$ and $z_2 = 100$, and we set $W_1 = W_2 = 1$. The background density of the plasma was $n_{\rm bkg} = 0.01$ so that $\omega_p^{\rm (bkg)} = 0.1$ and all the frequency content of the EM wave is above the plasma frequency of the background. We run 6 simulations with $n_{\rm max} = \{n_{\rm bkg}, 0.25, 0.5, 0.75, 1, 1.25\}$, that correspond to plasma frequencies at the top of the barrier $\omega_p^{\rm (max)} = \{0.1, 0.5, 0.707, 0.866, 1, 1.12\}$, respectively, and fall in different parts of the frequency spectrum of the EM wave packet. In the linear regime, we expect that the frequency components above $\omega_p^{\rm (max)}$ will propagate through the plasma barrier, while the others will be reflected, and this setup allows us to clearly appreciate how this mechanism takes place. 
In all these simulations we used a grid that extends in $[-1, 1] \times [-1, 1] \times [-450, 450]$, with a grid step $\Delta x = \Delta y = \Delta z = 0.2$ and a time step $\Delta t = 0.1$, so that the Courant-Friedrichs-Lewy factor is ${\rm CFL} = 0.5$ The final time of integration was set to $T = 400$.
\begin{figure*}[p]
	\centering
	\includegraphics[width = 0.85\textwidth]{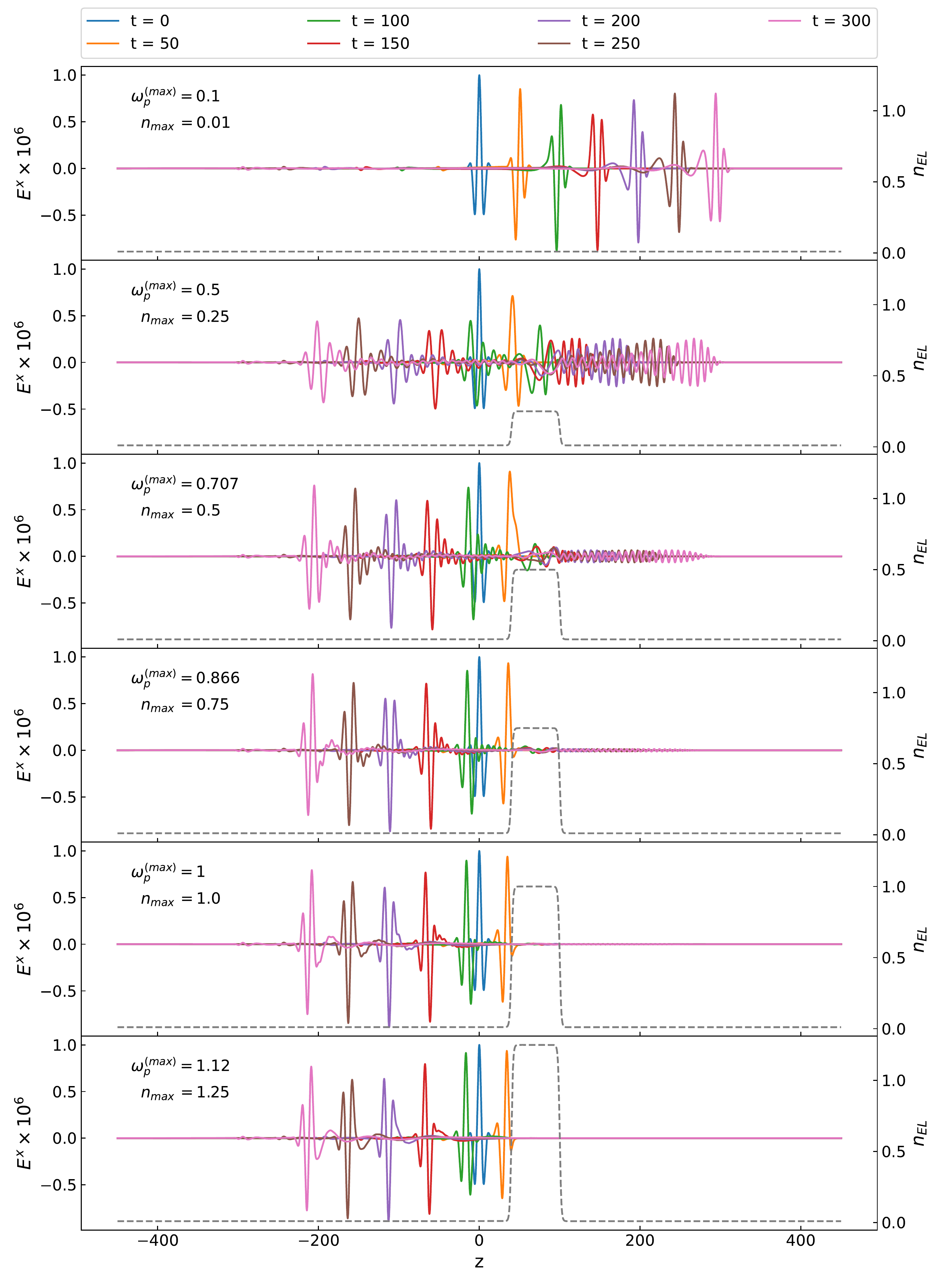}
	\caption{Propagation of an EM wave packet on a barrier of plasma in the linear regime. Here we show some snapshots of the evolution of $E^x$ for different values of the plasma density in the barrier, $n_{\rm max}$, and we represent the initial profile of $\nel$ with a gray dashed line.  When the plasma frequency in the barrier $\omega_p^{\rm (max)}$ becomes larger than $\omega$, the wave packet is mostly reflected by the barrier, while the transmitted component is suppressed. The corresponding animations are available online~\cite{webpage}.}
	\label{fig:LinearSnapshots}
\end{figure*}
Figure~\ref{fig:LinearSnapshots} shows some snapshots of the numerical results at different times for different values of $\omega_p^{\rm (max)}$. It is evident that the analytical predictions of linear theory are confirmed: as the plasma frequency of the barrier increases, less and less components are able to propagate through it and reach the other side. In particular, when $\omega_p^{\rm (max)}\gtrsim 0.9$ the wave is almost entirely reflected, and the transmitted component becomes negligible. As a matter of fact in this case, given the value of $\sigma$ that we set, on analytical grounds only a mere $\approx 2.5 \%$ of the initial wave packet spectrum lies above the plasma frequency and should penetrate through the barrier. Furthermore, in the linear regime the backreaction on the density is effectively negligible, as the barrier remains constant over the entire simulation (in fact, we observed a maximum variation of $\nel$ of the order of $10^{-11}$, which is clearly not appreciable on the scale of Fig.~\ref{fig:LinearSnapshots}).
% \enrico{in the linear case we can estimate as $n_{EL}=k_z E_z$? Is a bit trivial but could help w the referee}.

To better quantify the frequency components that are propagated and the agreement between the simulations and the analytic expectation in the linear regime, we computed the (discrete) Fourier transform of the time evolution of $E^x$ in two points along the $z$ axis: $z=-50$ and $z=150$, which are located before and after the plasma barrier, respectively. 
\begin{figure}
	\includegraphics[width = \columnwidth]{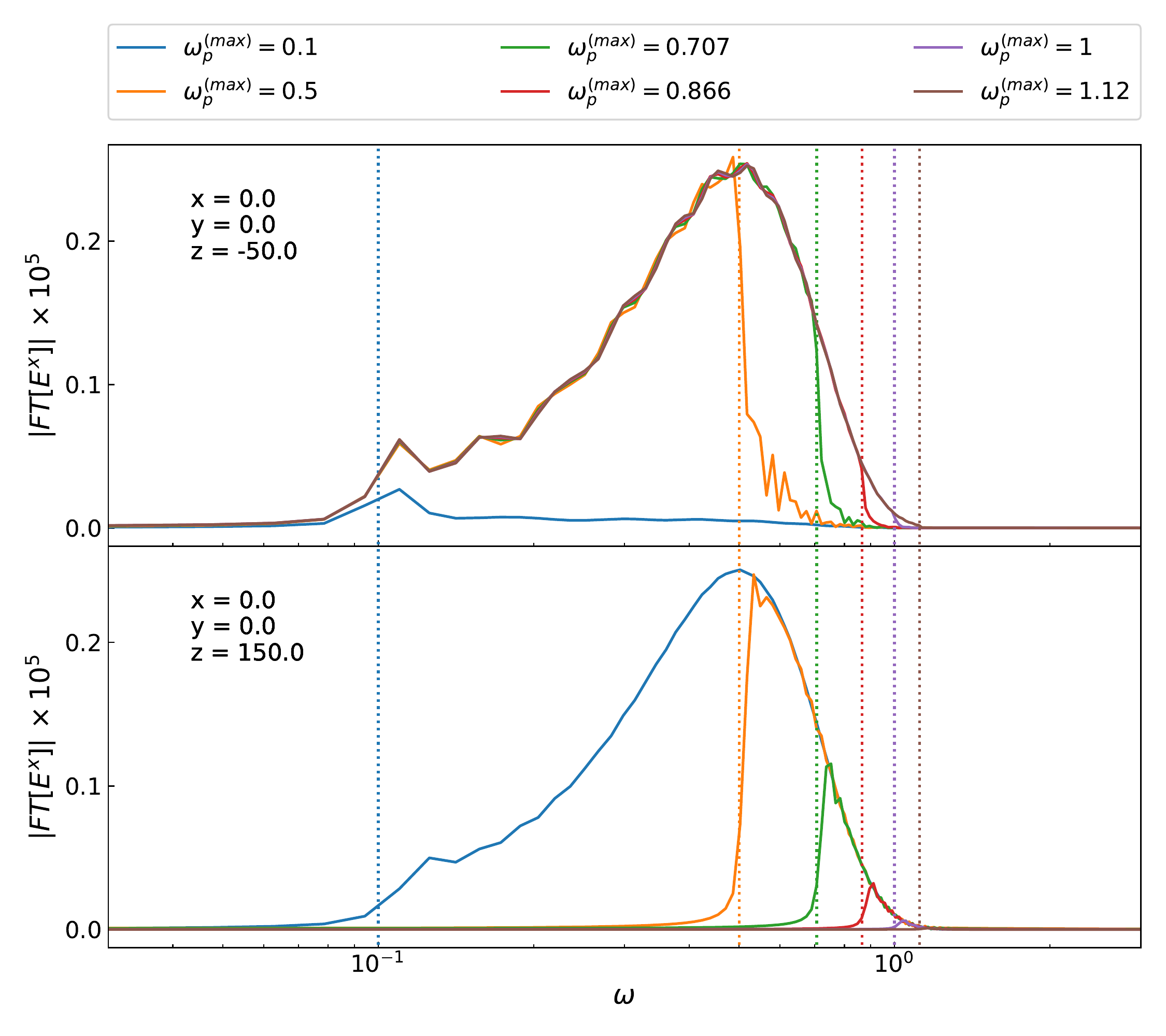}
	\caption{Absolute value of the discrete Fourier transform of $E^x$ extracted before (upper panel) and after (lower panel) the plasma barrier. The different colors refer to the different values of the plasma density inside the barrier, and hence to different values of the plasma frequency $\omega_p^{\rm (max)}$, indicated with vertical dotted lines. We can clearly see that the barrier reflects the frequency components below $\omega_p^{\rm (max)}$, and transmits the components above it.}
	\label{fig:LinearFourier}
\end{figure}
Figure~\ref{fig:LinearFourier} shows the absolute value of the Fourier transform for the different values of the plasma frequency in the barrier, which are represented as vertical dotted lines. As we can see from the Fourier transform at $z=150$, the transmitted waves have only components with frequency $\omega>\omega_p^{\rm (max)}$, in agreement with the fact that only modes above this threshold can propagate. Hence, the barrier perfectly acts as a high-pass filter, with a critical threshold given by the plasma frequency.

%%%%%%%%%%%%%
\subsection{Nonlinear regime}    \label{sec:results:nonlinear}
%%%%%%%%%%%%%
We can now proceed to increase the amplitude of the field until linear theory breaks down and the interaction becomes fully nonlinear. As anticipated, we shall show that the evolution is more involved than in the idealized model described in~\cite{1970PhFl...13..472K}. Indeed, even from a first qualitative analysis, it is evident from the $z$-component of the momentum equation~\eqref{eq:UUEvolCartesian} that in the nonlinear regime electrons will experience an acceleration along the $z$ axis due to the nonlinear Lorentz term $(\vec{\UU} \times \vec{B})^z$. The formation of a current along the $z$ directions implies a modification of the density profile because of the continuity equation, and also the formation of a longitudinal electric field that tries to balance and preserve charge neutrality. In the following, we will support this qualitative analysis with the results of the numerical simulations and show that nonlinear effects can have a dramatic impact on the system dynamics.

In this set of simulations, we set units\footnote{Note that, in rationalized Heaviside units, changing $m$ (and hence the classical electron radius) simply accounts for rescaling lengths, times, and masses in the simulations. Lengths and times are rescaled by $[m]^{-1}$, while the electric field amplitude scales as $[m]^2$. Hence, the results of this section can be obtained in the case $m=1$ by rescaling the other quantities accordingly.} such that $e=1$ and $m=1000$, and we consider an initial wave packet of the electric field centered at $z_0 = -150$, with a width\footnote{While formally the initial profile of the EM field, Eq.~\eqref{eq:EMInitial}, represents a circularly polarized wave packet, the chosen value of the parameter $\sigma$ reduces the $y$ component of the electric field, making the polarization effectively elliptic.} $\sigma = 100$ and $\omega=0.001$. We vary the amplitude of the EM in a range $0.1 \le A_E \le 1000$.  
As for the plasma profile, we adopt a similar geometric model to the linear case, with the barrier placed between $z_1 = 100$ and $z_2 = 650$, with $W_1 = W_2 = 0.1$. We consider a background density $n_{\rm bkg} = 5 \times 10^{-6}$, and a maximum barrier density $n_{\rm max} = 0.5$, that corresponds to a plasma frequency of $\omega_p^{\rm (max)} = 0.022$. We use a numerical grid that extends in $[-2, 2] \times [-2, 2] \times [-750, 850]$, with a grid spacing $\Delta x = \Delta y = \Delta z = 0.2$, and a time step $\Delta t = 0.1$, so that ${\rm CFL} = 0.5$. The final time of integration was set to $T = 500$.

The parameters are chosen such that the frequency of the wave packet is always much larger than $\omega_p^{\rm (bkg)}$, but a significant component of the spectrum, namely $\approx 97.5 \% $, is below the plasma frequency of the barrier, and should therefore be reflected if one assumes linear theory. 
\begin{figure*}[p]
	\centering
	\includegraphics[width = 0.85\textwidth]{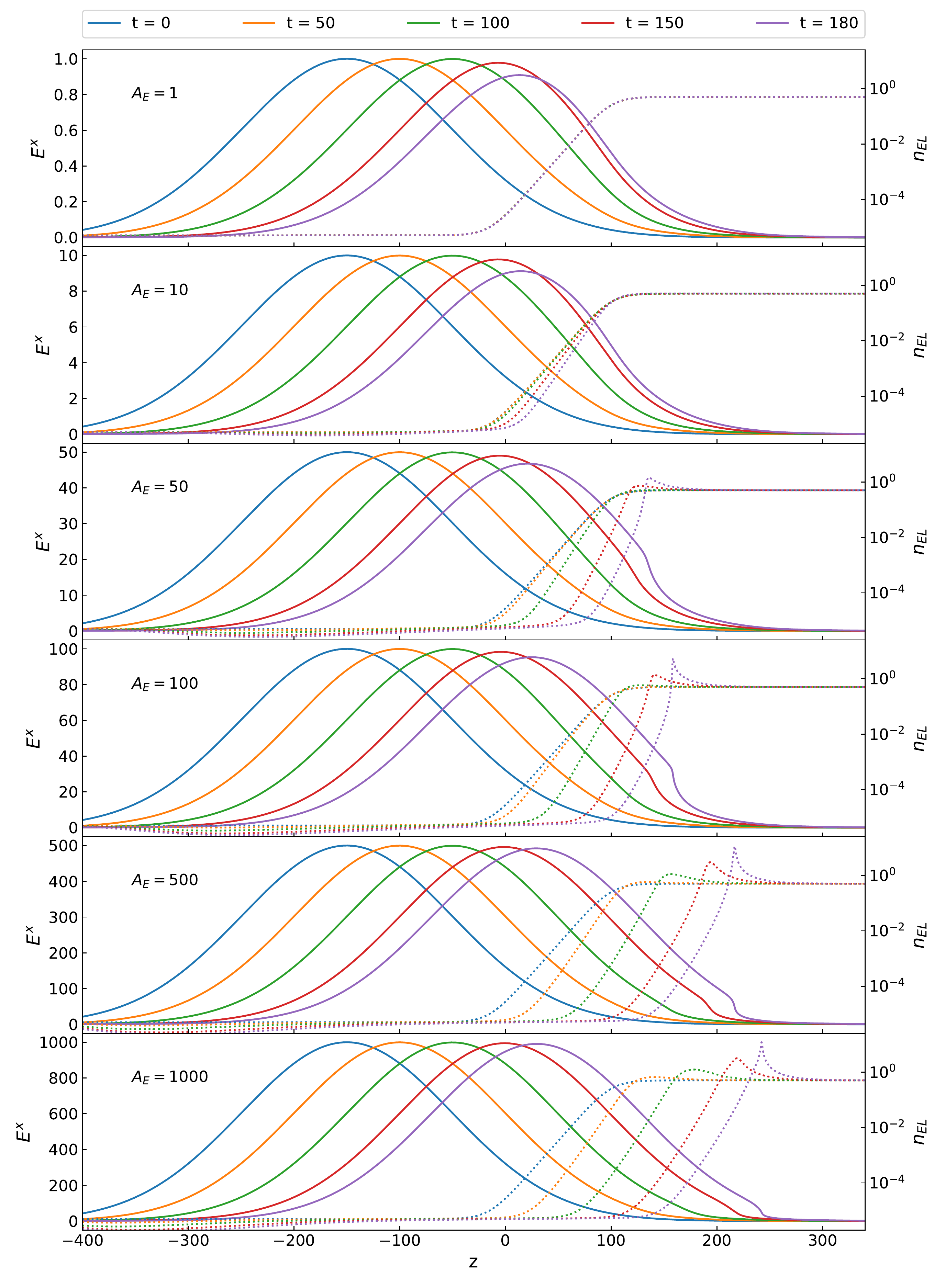}
	\caption{Snapshots of the evolution of $E^x$ (solid line) and $\nel$ (dashed lines) for the simulations of the propagation of an EM wave packet inside a plasma barrier in the nonlinear regime. The initial profile of $\nel$ is not varied across the simulations, and the different panels refer to different choices of the initial amplitude of the wave packet. The backreaction effects of the EM field onto the plasma density increase with $A_E$, and for $A_E \gtrsim 50$ wave packet ``transports'' electrons along the $z$ axis, eventually creating a plasma-depleted region (\textit{blowout regime}). The corresponding animations are available online~\cite{webpage}.}
	\label{fig:NonLinearSnapshots}
\end{figure*}
First of all, we quantify the value of the electric field which gives rise to nonlinearities. A crucial parameter that characterizes the threshold of nonlinearities in laser-plasma interactions is the peak amplitude of the normalized vector potential, defined as $a_0= e \textbf{A}/m$ (see e.g.~\cite{RevModPhys.81.1229,PhysRevLett.64.2011}).  Specifically, when $a_0 \gtrsim 1$, electrons acquire a relativistic transverse velocity, and therefore the interactions become nonlinear. Given our units, and estimating $A \approx E/\omega$, we obtain a critical electric field $E_{\rm crit} \gtrsim m \omega /e \approx 1$. 

We performed a set of simulations choosing different values of the initial amplitude of the EM wave packet in the range $0.1 \le A_E \le 1000$.
Figure~\ref{fig:NonLinearSnapshots} shows snapshots of the numerical simulations for some selected choices of $A_E$. It is possible to observe that in the case $A_E=1$ (top panel) the density profile of plasma is not altered throghout all the simulation, as in the linear case discussed in the previous section. Moreover, at sufficiently long times, the wavepacket is reflected by the barrier, in agreement with linear theory predictions.
From the second panel on (i.e. as $A_E \gtrsim 10$), instead, the wavepacket induces a nonnegligible backreaction on the plasma density. This effect increases significantly for higher amplitudes, and it is due to the nonlinear couplings between transverse and longitudinal polarizations: the nonlinear Lorentz term $(\vec{\UU} \times \vec{B})^z$ in the longitudinal component of the momentum equation~\eqref{eq:UUEvolCartesian} induces a radiation pressure on the plasma, and hence a longitudinal velocity $\vec{\UU}^z$; as electrons travel along the $z$ direction and ions remain at rest, a large longitudinal field due to charge separation is created, which tries to balance the effect of the Lorentz force and restore charge neutrality. This phenomenology resembles the one of plasma-based accelerators, where super-intense laser pulses are used to create large longitudinal fields that can be used to accelerate electrons~\cite{PhysRevLett.43.267}.

\begin{figure}
	\includegraphics[width = \columnwidth]{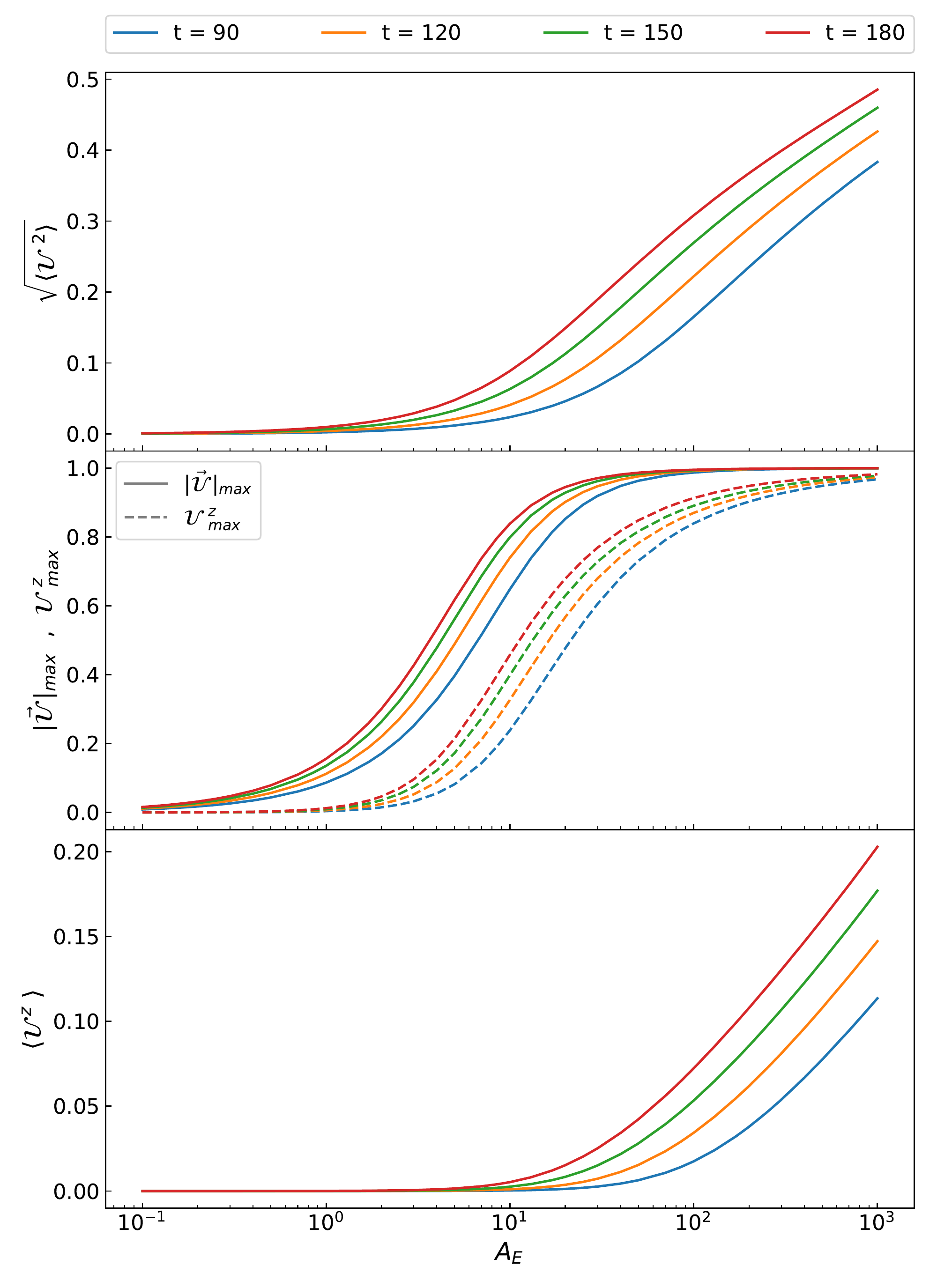}
	\caption{Collective behavior of plasma in the nonlinear regime as a function of the initial amplitude $A_E$ of the EM wavepacket. The upper panel shows the velocity dispersion $\sqrt{\langle \UU^2 \rangle}$, the middle panel shows the maximum value of $\lvert \vec \UU \rvert$ (solid lines) and of the longitudinal velocity $\UU^z$ (dashed lines), while the lower panel shows the collective longitudinal velocity $\langle \UU^z \rangle$. The nonlinearities start becoming relevant in the range $1 \lesssim A_E \lesssim 10$, where the velocity dispersion increases and the motion of electrons has a large component on the transverse plane. For $A_E \gtrsim 10$ the plasma enters in the blowout regime, where electrons are ``transported'' by the EM field, and acquire a positive collective longitudinal velocity.}	\label{fig:AverageVelocities}
\end{figure}

To quantify the collective motion induced by nonlinearities we computed the velocity dispersion of electrons as
\begin{equation}
    \sqrt{\langle \UU^2 \rangle} =\sqrt{\frac{\int_V d^3 x \,\nel \, \UU_i \UU^i}{\int_V \, d^3 x \, \nel}}.
    \label{eq:VelocityDispersion3D}
\end{equation}
Since the field are constant along the transverse directions\footnote{In Appendix~\ref{app:TransverseHomogeneity} we show how the homogeneity of the fields along the transverse plane is preserved during the evolution.}, then $\nel(x, y, z) = \nel(z)$ and $\UU^i(x, y, z) = \UU^i(z)$. This allows us to evaluate the above integral as
\begin{equation}
    \sqrt{\langle \UU^2 \rangle} =\sqrt{\frac{\int_{z_{-\infty}}^{z_{+\infty}} dz \, \nel(z) \, \UU_i(z) \UU^i(z)}{\int_{z_{-\infty}}^{z_{+\infty}} \, dz \, \nel(z)}},
    \label{eq:VelocityDispersion}
\end{equation}
where $z_{\pm\infty}$ are the boundaries of the $z$ domain and we compute the integral using the trapezoidal rule.  
In the upper panel of Fig.~\ref{fig:AverageVelocities} we plot the behavior of the velocity dispersion with respect to the initial amplitude $A_E$ for different times. As we can see the nonlinearities start becoming relevant in the range $1 \lesssim A_E \lesssim 10$, where electrons start to acquire a collective motion. This is also confirmed by the middle panel, where the solid and dashed lines denote the maximum of $\lvert \vec \UU \rvert$ and $\UU^z$, respectively. While these quantities do not represent the collective behavior of the system, they have the advantage of not containing the contribution given by the portion of the plasma barrier that has not been reached yet by the EM wave. From this plot we can observe that in the range $1 \lesssim A_E \lesssim 10$, the electrons start acquiring a relativistic velocity with a large component on the transverse plane.

As already mentioned, the longitudinal motion of electrons generate a longitudinal field. Nevertheless, plasmas can sustain longitudinal fields only up to a certain threshold, usually called \textit{wave-breaking (WB) limit}, above which plasma is not able to shield and sustain anymore electric fields, and the fluid description breaks down. This phenomenon was pioneered in~\cite{PhysRev.113.383} for the case of nonlinear, nonrelativistic cold plasmas, where the critical longitudinal field for WB was found to be $E^z_{\rm WB}=m\omega_p/e$, and later generalized for pulses with relativistic phase velocities~\cite{osti_4361348}. This threshold field represents the limit after which the plasma response loses coherence as neighbouring electrons start crossing each other within one plasma frequency period. Therefore, above this critical electric field the plasma is not anymore able to coherently act as a system of coupled oscillators, and the fluid model based on collective effects breaks down.  This leads to the formation of a spike in $\nel$, which eventually diverges, and to a steepening of the longitudinal component of the electric field. Full particle-in-cell numerical simulations are required after the breakdown (see, e.g.,~\cite{Pukhov:2002otp,PhysRevE.47.3585}). In our simulations, we observe the WB phenomenon at late time for large values of the electric field, in which cases we can only extract information before the breakdown of the model.

\begin{figure}
\includegraphics[width = \columnwidth]{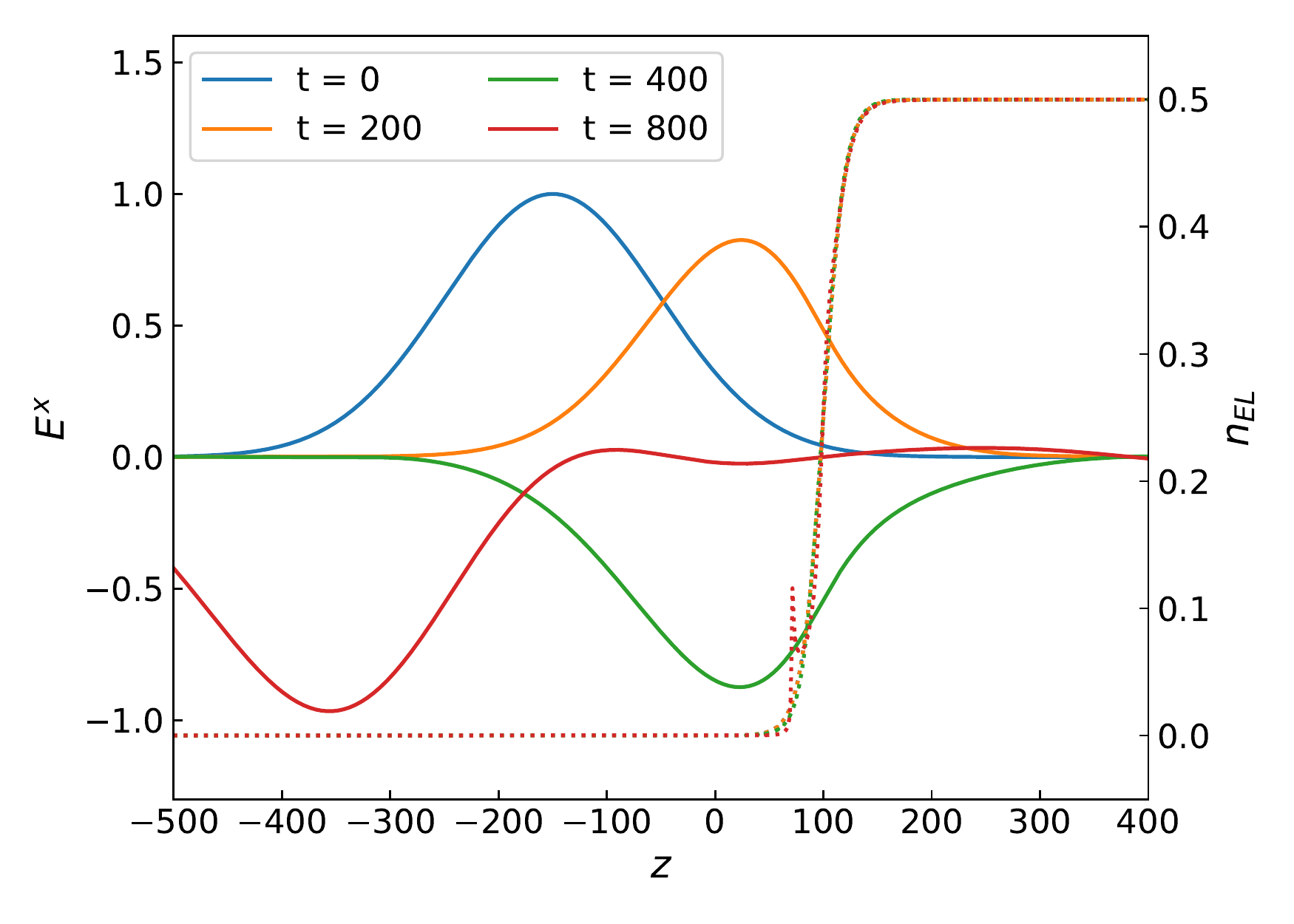}
\includegraphics[width = \columnwidth]{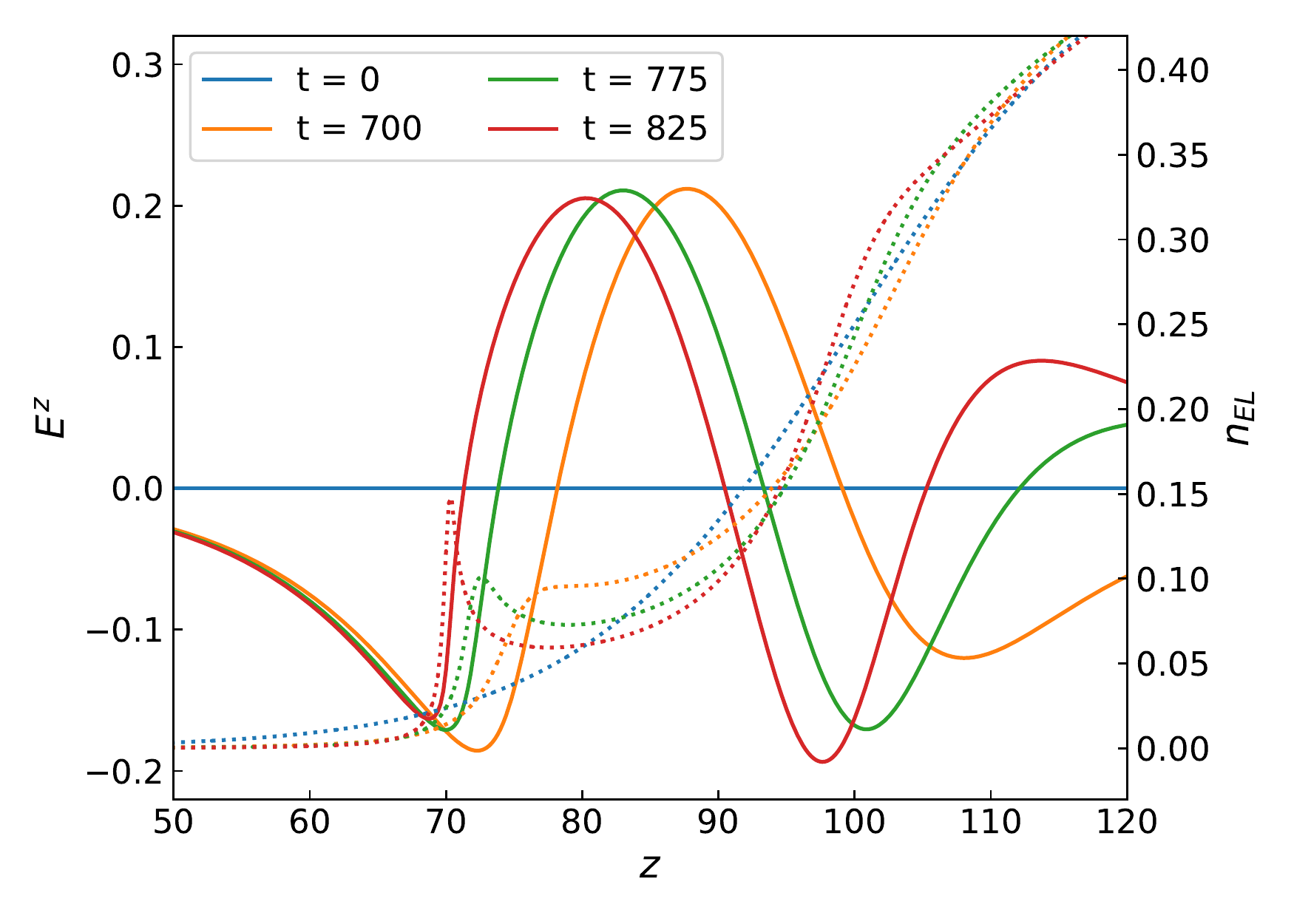}
\caption{ Evolution of the electric field and plasma density in the case of $A_E = 1$. Upper panel: snapshots of $E^x$ (solid lines) and $\nel$ (dotted lines) of the full evolution, where we can see that the wave packet is mostly reflected by the plasma barrier. Lower panel: snapshots of the longitudinal component $E^z$ (solid lines) and $\nel$ (dotted lines)  focusing on the last stages of evolution. Here we can clearly see the WB phenomenon taking place, with the plasma density developing spikes in regions where the longitudinal component of the electric field increases steeply.}
\label{fig:WaveBreakingAtTheThreshold}
\end{figure}

In order to better appreciate how the WB takes place, we repeated the simulation with $A_E = 1$ for a longer integration time and a larger grid.
In the upper panel of Fig.~\ref{fig:WaveBreakingAtTheThreshold} we show the evolution of $E^x$ (solid lines) throughout all the simulation, where we can clearly see that the incoming wave packet is reflected by the plasma barrier. However, for $t \approx 700$, the longitudinal component of $\vec \UU$ leads to an evolution of the plasma density. In this stage the plasma loses coherence and $\nel$ develops local spikes that increase in height and becomes sharper with time. When one of these spikes becomes excessively narrow, the fluid description of the system breaks down and the simulation crashes. This can be observed from the bottom panel of Fig.~\ref{fig:WaveBreakingAtTheThreshold}, where we show the longitudinal component of $\vec E$ together with the plasma density profile. Note that WB occurs as soon as the nonlinearities come into play (we observed it already for $A_E = 1$), and the fluid description in the nonlinear regime cannot be used for long-term numerical simulations. However the good convergence of the code even slightly before WB takes place (see Appendix~\ref{app:Convergence}) ensures the reliability of the results up to this point.
Finally, notice that one can obtain an analytical estimate of the critical WB field and achieve a good agreement with the numerical simulation. In particular, from Fig.~\ref{fig:WaveBreakingAtTheThreshold} we estimated the local plasma density at which WB takes place to be $n_{EL}\approx 0.01$, so that the critical longitudinal field is $E^z_{\rm WB}=m\omega_p/e \approx 3$, which is comparable with $A_E=1$ used in Fig.~\ref{fig:WaveBreakingAtTheThreshold}.

Overall, Figs. \ref{fig:NonLinearSnapshots} and \ref{fig:WaveBreakingAtTheThreshold} show that for $A_E\sim 1$ the system becomes weakly nonlinear, in agreement with the previously mentioned analytical estimates.

Going back to the snapshots of the evolutions in Fig.~\ref{fig:NonLinearSnapshots}, we now wish to analyze the behavior of the system for larger electric fields, where the backreaction is macroscopic. We can see that in this case, i.e. for $A_E \gtrsim 50$, all the electrons in the plasma barrier are ``transported'' in the $z$ direction and piled up within a plasma wake whose density grows over time. This corresponds to a \textit{blowout regime} induced by radiation pressure. In order to better describe how the system reaches this phase, we can compute the longitudinal component of the collective electron velocity as
\begin{equation}
    \langle \UU^z \rangle = \frac{\int_V d^3 x \, \nel \, \UU^z}{\int_V d^3 x \, \nel} = \frac{\int_{z_{-\infty}}^{z_{+\infty}} dz \, \nel(z) \, \UU^z(z)}{\int_{z_{-\infty}}^{z_{+\infty}} \, dz \, \nel(z)},
    \label{eq:AverageVz}
\end{equation}
where, again, we took advantage of the homogeneity of the system along the transverse direction to reduce the dimensionality of the domain of integration. 
The results are shown in the lower panel of Fig.~\ref{fig:AverageVelocities}, where we can see that for $A_E \lesssim 10$, the longitudinal momentum remains low and is not influenced by the wavepacket. For $A_E \gtrsim 10$ instead, $\langle \UU^z \rangle$ starts to increase in time, indicating that the system is in the blowout regime, as electrons are collectively moving forward in the $z$ direction.

Overall, the above analysis shows that when the idealized situation studied in~\cite{1970PhFl...13..472K} cannot be applied and the nonlinear Lorentz term does not vanish, the general physical picture is drastically different and that penetration occurs in this setup due to radiation-pressure acceleration rather than transparency.

%%%%%%%%%%%%%%%%%%%%%%%%%%%%%%
\section{Discussion: Implications for plasma-driven superradiant instabilities}
%%%%%%%%%%%%%%%%%%%%%%%%%%%%%%
Motivated by exploring the plasma-driven superradiant instability of accreting BHs at the full nonlinear level, we have performed $3+1$ numerical simulations of a plane wave of very large amplitude but small frequency scattered off an inhomogeneous plasma barrier.
Although nonlinear plasma-photon interactions are well studied in plasma-physics applications, to the best of our knowledge this is the first analysis aimed at exploring numerically this interesting setup in generic settings.

One of our main findings is the absence of the relativistic transparency effect in our simulations. As already mentioned, the analysis performed in~\cite{1970PhFl...13..472K} showed that, above a critical electric field, plasma turns from opaque to transparent, thus enabling the propagation of EM waves with frequency below the plasma one. From Eq.~\eqref{omegapnonlin}, such critical electric field for transparency is $E_{\rm crit}^{\rm transp}= \frac{m}{e} \sqrt{\frac{\omega_p^4}{\omega^2} - \omega^2}$. In our simulations, we considered electric fields well above this threshold, yet we were not able to observe this effect. On the contrary, in the nonlinear regime the plasma strongly interacts with the EM field in a complex way.
The role of relativistic transparency in more realistic situations than the one described in~\cite{1970PhFl...13..472K} was rarely considered in the literature and is still an open problem~\cite{2016APS..DPPTO6007S}. Nevertheless, some subsequent analysis found a number of interesting features, and revealed that its phenomenology in realistic setups is more complex.

In Ref.~\cite{Cattani} an analytical investigation of a similar setup was performed by considering the scattering between a laser wavepacket and a sharp boundary plasma. The conclusion of the analysis is that, when plasma is inhomogeneous, nonlinearities tend to create a strong peaking of the plasma electron density (and hence of the effective plasma frequency), suppressing the laser penetration and enhancing the critical threshold needed for transparency. 
Subsequently, Refs.~\cite{PhysRevLett.87.275002, Berezhiani:2005} confirmed this prediction numerically, and showed that in a more realistic scenario transparency can occur but the phenomenology is drastically different from the one predicted in~\cite{1970PhFl...13..472K}. For nearly-critical plasmas, transparency arises due to the propagation of solitons, while for higher densities the penetration effect holds only for finite length scales. Nevertheless, these simulations were performed by considering a simplified momentum equation due to the assumption of a null-vorticity plasma, which is typically suitable for unidimensional problems, but likely fails to describe complex-geometry problems as the one of superradiant fields. Using particle-in-cell simulations, it was then realized that radiation-pressure can push and accelerate the fluid to relativistic regimes, similarly to our results, and produce interesting effects such as hole-boring, ion acceleration, and light-sail~\cite{PhysRevLett.79.2686, macchi_2014}.

\begin{figure}
    \includegraphics[width = \columnwidth]{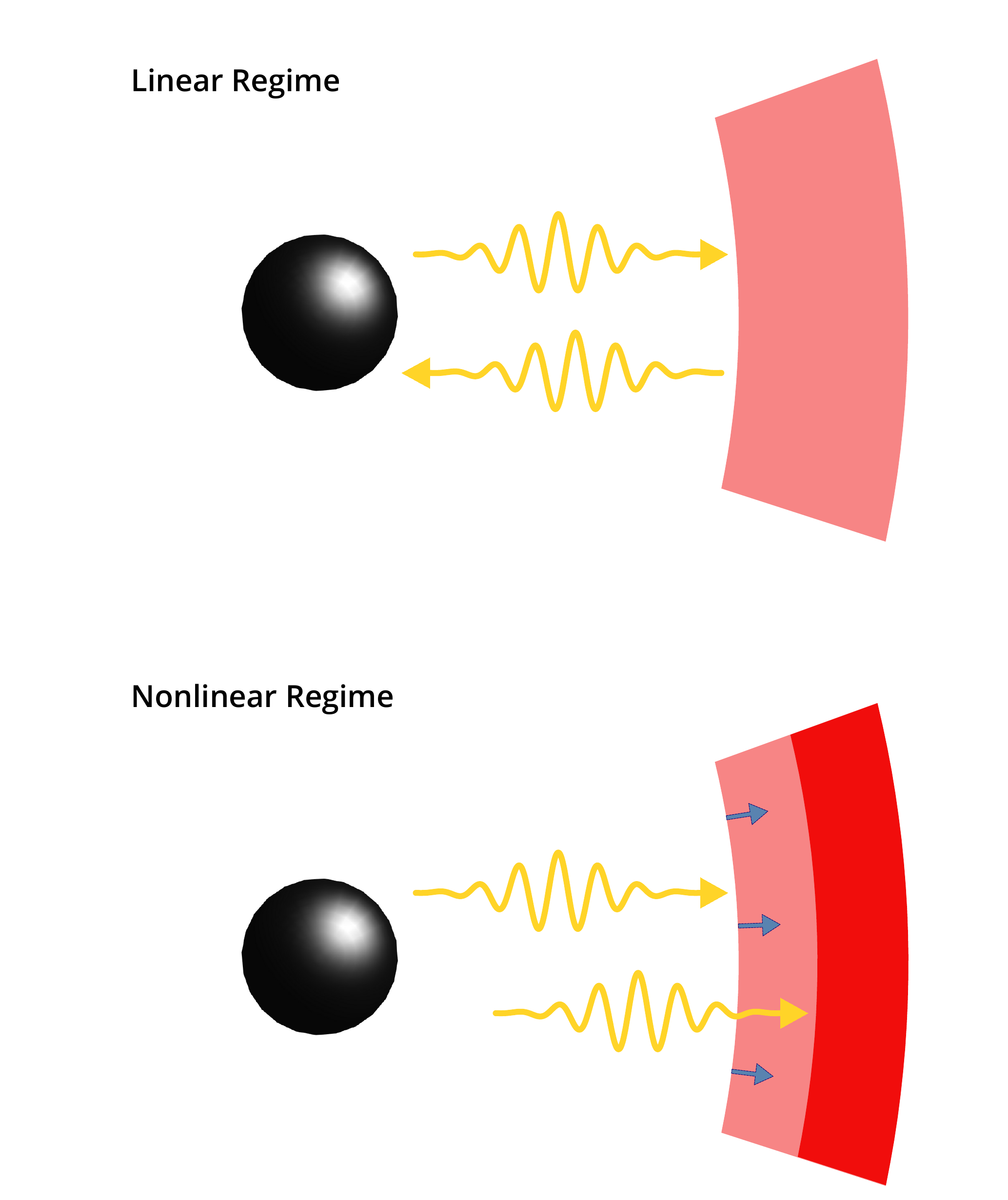}
    \caption{Pictorial representation of the photon-plasma interaction in the context of BH superradiance. While in the linear regime the plasma profile does not evolve and is able to efficiently act as a mirror, when the EM field grows and the system enters in the nonlinear regime, the radiation pressure pushes the plasma away and jeopardizes the instability.}
    \label{fig:PhenomenologicalPicture}
\end{figure}
While the complicated interplay between relativistic transparency and radiation-pressure acceleration is still an open problem~\cite{2016APS..DPPTO6007S, Bulanov_2016}, we argue that the latter, which arises in generic situations with very overdense plasmas and high amplitude electric fields, is sufficient to dramatically quench
the plasma-driven superradiant instability. 
Indeed, in order to have an efficient instability akin to the BH bomb, the plasma should be able to reflect EM radiation as a perfect mirror. While our simulations feature this behavior in the linear regime, in the nonlinear case the large EM field exerts a pressure on the plasma profile, pushing it away and possibly rendering the confinement ineffective. A pictorial representation of this phenomenology is shown in Fig.~\ref{fig:PhenomenologicalPicture}.
To enforce this conclusion, we provide a rough estimate of the total energy extracted from the BH before nonlinear effects take place~\cite{Cardoso:2020nst}. 
In order for the instability to be efficient on astrophysical timescales, $\omega\lesssim\omega_p \approx O(1/(GM))$, where $G$ is Newton's constant and $M$ is the BH mass~\cite{Dima:2020rzg,Cannizzaro:2020uap,Cannizzaro:2021zbp}. This gives a critical electric field
\begin{equation}
    E_{\rm crit}=\frac{m \omega}{e}\approx 4 \times 10^5 \frac{\rm V}{\rm cm}\Big(\frac{M_\odot}{M} \Big) 
\end{equation}
The associated total energy can be estimated as $U= E_{\rm crit}^2 L^3$, where $L$ is the size of the condensate formed by the superradiant instability, and corresponds to the location of the plasma barrier. This gives
\begin{equation}
U\approx 10^7 {\rm J}\Big(\frac{M}{M_\odot} \Big)\Big(\frac{L}{6M} \Big)^3 \,
\end{equation}
where we assumed that the peak of the plasma barrier roughly corresponds to the location of the peak density of an accretion disk, $L\approx 6 M$. On the other hand,  the total rotational energy of the BH is given by $K= M R^2 \Omega^2$, where $R$ and $\Omega$ are the radius and the angular velocity of the horizon, respectively. To efficiently satisfy the superradiant condition, $\Omega\gtrsim\omega_p\approx {\cal O}(1/(GM))$, so that
\begin{equation}
    K\approx 10^{43} {\rm J} \Big(\frac{M}{M_\odot} \Big)\,.
\end{equation}
Therefore, when the electric field reaches the threshold for nonlinearities, the total energy extracted from the BH is tiny, $U/K \approx 10^{-36}$.

Another argument supporting this conclusion is that, for the superradiant instability to be sustainable, the maximum energy leakage of the confining mechanism cannot exceed the superradiant amplification factor of the BH. For EM waves, the maximum amplification factor (for nearly extremal BHs and fine-tuned frequency) does not exceed $\approx 4\%$ and is typically much smaller~\cite{Brito:2015oca}. Therefore, the instability is not quenched only if the plasma is able to confine more than $96\%$ of the EM field energy. Our simulations shows that in the nonlinear regime the situation is quite the opposite: almost the entirety of the EM field is not confined by the plasma, thus destroying its capability to ignite the instability.
We expect this argument to be valid also when $\omega_p\gg\omega$, in which case plasma depletion through blowout is negligible, but the EM field can still transfer energy into longitudinal plasma motion. 

Note that the arguments above are extremely conservative, since are based on a number of optimistic assumptions that would maximize the instability. First of all, realistic accretion flows around BHs are not spherical nor stationary, especially around spinning BHs. This would generically introduce mode-mixing and decoherence, rendering the instability less efficient. More importantly, even in the linear regime a disk-shape accretion geometry can (partially) confine modes that are mostly distributed along the equatorial plane, but would naturally provide energy leakage along off-equatorial directions~\cite{Lingetti:2022psy,Wang:2022hra}. Finally, a sufficiently high plasma density in the corona could quench photon propagation in the first place~\cite{Dima:2020rzg}, at least at the linear level during the early stages of the instability.

Although our results strongly suggest that nonlinearities completely quench the ordinary plasma-triggered BH superradiant instability, our framework can be directly used to explore more promising problems in other contexts, especially in beyond-Standard-Model scenarios.
It would be interesting to study how nonlinear plasma interactions affects BH superradiant instabilities triggered by ultralight bosons, for example in the context of axion electrodynamics or in the case of superradiant dark photons kinetically mixed with ordinary photons.
In the latter case, if the plasma frequency is much greater than the dark photon bare mass, the two vector fields decouple due to in-medium suppressions. In Refs.~\cite{Caputo:2021efm,Siemonsen:2022ivj}, it was assumed that as the dark photon field grows and accelerates the plasma, the effect of the plasma frequency vanishes as it is unable to impede the propagation of high amplitude EM waves. While our results confirm these statements, they also prove that in generic settings the propagation will dramatically alter the plasma profile, and therefore suggest that even in these systems (as well as for axion-photon induced blasts~\cite{Rosa:2017ury,Ikeda:2018nhb,Boskovic:2018lkj, Spieksma:2023vwl}) a more careful analysis at the plasma frequency scale must be performed. 

\begin{acknowledgments}
We thank Andrea Caputo for interesting conversations. We acknowledge financial support provided under the European
Union's H2020 ERC, Starting Grant agreement no.~DarkGRA--757480 and support under the MIUR PRIN (Grant 2020KR4KN2 “String Theory as a bridge between Gauge Theories and Quantum Gravity”) and FARE (GW-NEXT, CUP: B84I20000100001, 2020KR4KN2) programmes.
We also acknowledge additional financial support provided by Sapienza, ``Progetti per Avvio alla Ricerca'', protocol number AR1221816BB60BDE.
\end{acknowledgments}

\appendix
\section{Derivation of the $3+1$ form of the field equations} \label{app:3+1}

Here we perform the explicit computation to obtain the field equations in the $3+1$ form. For the EM field we avoid to rewrite the procedure and we refer directly to~\cite{Alcubierre:2009ij}. We will thus consider only Eqs.~\eqref{eq:FieldPlasmaEvolution}, \eqref{eq:FieldPlasmaContinuity}.

\subsection{Decomposition of Eq.~\eqref{eq:FieldPlasmaEvolution}}

Let us rewrite Eq.~\eqref{eq:FieldPlasmaEvolution} for clarity:
\begin{equation}
	u^\nu \nabla_\nu u^\mu = \frac{e}{m} F^{\mu\nu} u_\nu.	
\end{equation}
we have to project it separately on $n^\mu$ and on $\Sigma_t$.

\subsubsection*{Projection on $n^\mu$}

Contracting Eq.~\eqref{eq:FieldPlasmaEvolution} with $n_\mu$ we obtain
\begin{equation}
	n_\mu u^\nu \nabla_\nu u^\mu = \frac{e}{m} F^{\mu\nu} u_\nu n_\mu.	
	\label{eq:FPEprojectionEul}
\end{equation}
In the right hand side we have
\begin{equation}
	\frac{e}{m} F^{\mu\nu} u_\nu n_\mu = - \frac{e}{m} E^{\nu} u_\nu = - \frac{e}{m} E^{\nu} \spatial{u}_\nu,
	\label{eq:FPEprojectionEulRHS}
\end{equation}
where in the last step we used the fact that $E^\mu$ lies on $\Sigma_t$. 
The left hand side requires more manipulation. In particular we have that
\begin{align}
	n_\mu u^\nu \nabla_\nu u^\mu &= u^\nu \nabla_\nu (n_\mu u^\mu) - u^\mu u^\nu \nabla_\nu n_\mu \nonumber \\
                                 &= -u^\nu \nabla_\nu \Gamma - u^\mu u^\nu \nabla_\nu n_\mu.
	\label{eq:FPEprojectionEulLHS}
\end{align}
Let us now consider only the second term:
\begin{align}
	u^\mu u^\nu \nabla_\nu n_\mu &= u^\mu u^\nu \tensor{\delta}{^\lambda_\nu} \nabla_\lambda n_\mu = u^\mu u^\nu (\sproj{^\lambda_\nu}  - n^\lambda n_\nu) \nabla_\lambda n_\mu \nonumber \\
 &= u^\nu u^\mu \sproj{^\lambda_\nu} \nabla_\lambda n_\mu - u^\nu n_\nu u^\mu a_\mu \notag \\
 &=  u^\nu u^\mu \sproj{^\lambda_\nu} \tensor{\delta}{^\sigma_\mu} \nabla_\lambda n_\sigma + \Gamma u^\mu a_\mu \nonumber \\
 &= u^\nu u^\mu \sproj{^\lambda_\nu} \sproj{^\sigma_\mu} \nabla_\lambda n_\sigma - u^\nu u^\mu \sproj{^\lambda_\nu} n^\sigma n_\mu \nabla_\lambda n_\sigma \nonumber \\
 &+ \Gamma u^\mu a_\mu.
\end{align}
Here we used the definition of the projection operator $\sproj{^\mu_\nu} = \tensor{\delta}{^\mu_\nu} + n^\mu n_\nu$, the definition of $\Gamma$, and defined the 4-acceleration of the Eulerian observer, $a_\mu = n^\nu \nabla_\nu n_\mu = D_\mu \ln\alpha$. 
Given that $n^\mu n_\mu = -1$ the second term in the last line vanishes. 
Furthermore, by recognizing that $K_{\mu\nu} = -\sproj{^\lambda_\nu} \sproj{^\sigma_\mu} \nabla_\lambda n_\sigma$, we can write the first term as $- K_{\mu\nu} u^\mu u^\nu$. Substituting all these terms in Eq.~\eqref{eq:FPEprojectionEul} we obtain
\begin{equation}
	-u^\mu \nabla_\mu \Gamma + K_{\mu\nu} u^\mu u^\nu - \Gamma u^\mu D_\mu \ln{\alpha} = - \frac{e}{m} E^{\mu} \spatial{u}_\mu.
\end{equation}
Using now the decomposed form of $u^\mu$ (Eq.~\eqref{eq:uDecomposition}) we can write
\begin{align}
	\tder \Gamma &= \beta^i \partial_i \Gamma - \alpha \UU^i \partial_i \Gamma + \alpha \Gamma K_{ij} \UU^i \UU^j \nonumber \\
                 &- \Gamma \UU^i \partial_i \alpha + \frac{e}{m} \alpha E^i \UU_i.
	\label{eq:GammaEvolGeneric:app}
\end{align}

\subsubsection*{Projection on $\Sigma_t$}

Let us now project Eq.~\eqref{eq:FieldPlasmaEvolution} with $\sproj{^\mu_\nu}$:
\begin{equation}
	\sproj{^\mu_\sigma}u^\nu \nabla_\nu u^\sigma = \frac{e}{m} \sproj{^\mu_\sigma} F^{\sigma\nu} u_\nu.
	\label{eq:FPEProjectionSigma}
\end{equation}
In the right hand side we have
\begin{align}
	\frac{e}{m} \sproj{^\mu_\sigma} F^{\sigma\nu} u_\nu &= \frac{e}{m} \sproj{^\mu_\sigma}(n^\sigma E^\nu - n^\nu E^\sigma + \tensor{\spatial{\epsilon}}{^{\sigma\nu\lambda}}B_\lambda) u_\nu \nonumber  \\
	&= - \frac{e}{m} n^\nu u_\nu E^\mu + \frac{e}{m} \tensor{\spatial{\epsilon}}{^{\mu\nu\lambda}}B_\lambda u_\nu \nonumber \\
	&= \frac{e}{m} \Gamma E^\mu + \frac{e}{m} \Gamma \tensor{\spatial{\epsilon}}{^{\mu\nu\lambda}}B_\lambda \UU_\nu.
	\label{eq:FPEProjectionSigmaRHS}
\end{align}
In the left hand side, instead, we start by substituting the decomposition \eqref{eq:uDecomposition}:
\begin{widetext}
    \begin{align}
    	\sproj{^\mu_\sigma}u^\nu \nabla_\nu u^\sigma &= \sproj{^\mu_\sigma}u^\nu \nabla_\nu (\Gamma n^\sigma + \spatial{u}^\sigma) = \sproj{^\mu_\sigma} u^\nu n^\sigma \nabla_\nu \Gamma + \Gamma \sproj{^\mu_\sigma} u^\nu \nabla_\nu n^\sigma + \sproj{^\mu_\sigma} u^\nu \nabla_\nu \spatial{u}^\sigma \notag \\
    	&= \Gamma \sproj{^\mu_\sigma} (\Gamma n^\nu + \spatial{u}^\nu) \nabla_\nu n^\sigma + \sproj{^\mu_\sigma}(\Gamma n^\nu + \spatial{u}^\nu) \nabla_\nu \spatial{u}^\sigma \notag \\
    	&= \Gamma^2 \sproj{^\mu_\sigma} a^\sigma - \Gamma \tensor{K}{^\mu_\nu}\spatial{u}^\nu + \Gamma \sproj{^\mu_\sigma} n^\nu \nabla_\nu \spatial{u}^\sigma + \sproj{^\mu_\sigma} \spatial{u}^\nu D_\nu \spatial{u}^\sigma,
    	\label{eq:FPEProjectionSigmaLHSStep1}
    \end{align}
\end{widetext}
where in the third step we used the orthogonality between $n^\mu$ and $\sproj{^\mu_\nu}$, while on the fourth step we used the definition of the 4-acceleration $a_\mu$ and the extrinsic curvature $K_{\mu\nu}$. The covariant derivative $D_\mu$ has been introduced according to the definition $D_\nu \spatial{u}^\mu = \sproj{^\sigma_\nu} \sproj{^\mu_\lambda} \nabla_\sigma \spatial{u}^\lambda$. 
Let us now rewrite this equation in terms of $\UU^\mu$:
\begin{align}
	\sproj{^\mu_\sigma}u^\nu \nabla_\nu u^\sigma &= \Gamma^2 a^\mu - \Gamma^2 \tensor{K}{^\mu_\nu} \UU^\nu + \Gamma \UU^\mu n^\nu \nabla_\nu \Gamma \nonumber \\
                                                 &+ \Gamma^2 \sproj{^\mu_\sigma} n^\nu \nabla_\nu \UU^\sigma + \UU^\nu \UU^\mu \Gamma D_\nu \Gamma \nonumber \\
                                                 &+ \Gamma^2 \sproj{^\mu_\sigma} \UU^\nu D_\nu \UU^\sigma.
	\label{eq:FPEProjectionSigmaLHSStep2}
\end{align}

Now we wish to rewrite the spatial components of this equation in the form of an evolution equation, and for this purpose we use a procedure similar to the one in Eqs.~(A14) - (A20) of~\cite{Alcubierre:2009ij}. 
First we note that for any 3-vector $\spatial{V}^\mu$, $\LL_n \spatial{V}^\nu = n^\mu \nabla_\mu \spatial{V}^\nu - \spatial{V}^\mu \nabla_\mu n^\nu$, so that
\begin{align}
	\sproj{^\nu_\sigma} n^\mu \nabla_\mu \spatial{V}^\sigma &= \sproj{^\nu_\sigma} \LL_n \spatial{V}^\sigma + \sproj{^\nu_\sigma} \spatial{V}^\mu \nabla_\mu n^\sigma \nonumber \\
                                    &= \sproj{^\nu_\sigma} \LL_n \spatial{V}^\sigma - \spatial{V}^\mu \tensor{K}{^\nu_\mu}.
	\label{eq:nabla_n_3vec}
\end{align}
Now, the Lie derivative can also be written in terms of partial derivatives, and setting $\nu = i$ we obtain 
\begin{align}
    \sproj{^i_\sigma} n^\mu \nabla_\mu \spatial{V}^\sigma &= \sproj{^i_\sigma} \LL_n \spatial{V}^\sigma - \spatial{V}^j \tensor{K}{^i_j} \nonumber \\
                                                          &= \frac{1}{\alpha} \tder \spatial{V}^i - \frac{\beta^j}{\alpha} \partial_j \spatial{V}^i + \frac{\spatial{V}^j}{\alpha} \partial_j \beta^i \nonumber \\
                                                          &- \spatial{V}^j \tensor{K}{^i_j}\,,
    \label{eq:nabla3vec_partial_derivatives}
\end{align}
where we made use of the explicit expressions of $\sproj{^\mu_\nu}$ and $n^\mu$.

If we now substitute Eq.~\eqref{eq:nabla3vec_partial_derivatives} in the $i$-th component of Eq.~\eqref{eq:FPEProjectionSigmaLHSStep2}, we get
\begin{align}
    \sproj{^i_\sigma}u^\nu \nabla_\nu u^\sigma &= \Gamma^2 a^i + \Gamma \UU^i n^\nu \nabla_\nu \Gamma + \UU^i \UU^j \Gamma D_j \Gamma \nonumber \\
                                               &+ \frac{\Gamma^2}{\alpha} \Bigl( \tder \UU^i - \beta^j \partial_j \UU^i + \UU^j \partial_j \beta^i \Bigr) \nonumber \\
                                               &+ \Gamma^2 \UU^j D_j \UU^i - 2\Gamma^2 \tensor{K}{^i_j} \UU^j\,.
	\label{eq:FPEProjectionSigmaLHSStep3}
\end{align}
Next, $n^\nu \nabla_\nu \Gamma = \frac{1}{\alpha} [\tder \Gamma - \beta^i \partial_i \Gamma]$, which is given by Eq.~\eqref{eq:GammaEvolGeneric:app}. Substituting in Eq.~\eqref{eq:FPEProjectionSigmaLHSStep3} we obtain
\begin{align}
    \sproj{^i_\sigma}u^\nu \nabla_\nu u^\sigma &= \Gamma^2 a^i + \Gamma^2 \UU^i K_{jl} \UU^j \UU^l - \Gamma^2 \UU^i \UU^j \frac{\partial_ j \alpha}{\alpha} \nonumber \\
                                               &+ \frac{\Gamma^2}{\alpha} \Bigl( \tder \UU^i - \beta^j \partial_j \UU^i + \UU^j \partial_j \beta^i \Bigr) \nonumber \\
                                               &+ \frac{e}{m} \Gamma \UU^i E^j \UU_j + \Gamma^2 \UU^j D_j \UU^i - 2\Gamma^2 \tensor{K}{^i_j} \UU^j\,.                                        
	\label{eq:FPEProjectionSigmaLHS}
\end{align}

We are now ready to replace Eq.~\eqref{eq:FPEProjectionSigmaLHS} and the spatial components of Eq.~\eqref{eq:FPEProjectionSigmaRHS} in the original equation~\eqref{eq:FPEProjectionSigma} and isolate the evolution operator. The result is:
\begin{align}
    \tder \UU^i &= \beta^j \partial_j \UU^i - \UU^j \partial_j \beta^i - \alpha a^i - \alpha \UU^i K_{jl} \UU^j \UU^l \nonumber \\
                &+ \frac{\alpha}{\Gamma} \frac{e}{m} \Bigl( - \UU^i E^j \UU_j + E^i + \tensor{\spatial{\epsilon}}{^{i j l}} B_l \UU_j \Bigr) \nonumber \\
                &+ 2 \alpha \tensor{K}{^i_j} \UU^j + \UU^i \UU^j \partial_j \alpha - \alpha \UU^j D_j \UU^i\,.
                \label{eq:UUEvolGeneric:app}
\end{align}

\subsection{Continuity equation in $3+1$ variables}

Let us now use the variables that we have introduced to rewrite the continuity equation Eq.~\eqref{eq:FieldPlasmaContinuity}.
Using the decomposition $u^\mu = \Gamma (n^\mu + \UU^\mu)$ and the definition of the electron density seen by the Eulerian observer, $\nel = \Gamma n_e$, we can rewrite Eq.~\eqref{eq:FieldPlasmaContinuity} as 
\begin{align}
	0 &= \nabla_\mu [n_e \Gamma (n^\mu + \UU^\mu)] = \nabla_\mu [\nel (n^\mu + \UU^\mu)] \nonumber \\
      & =  n^\mu \nabla_\mu \nel + \UU^\mu \nabla_\mu \nel + \nel \nabla_\mu n^\mu + \nel \nabla_\mu \UU^\mu\,.
	\label{eq:PFCDecompositionGeneric}
\end{align}
Expressing $n^\mu \nabla_\mu \nel$ in terms of Lie derivatives, Eq.~\eqref{eq:PFCDecompositionGeneric} can be written as an evolution equation for $\nel$:
\begin{equation}
	\tder \nel = \beta^i \partial_i \nel  + \alpha K \nel  - \alpha \UU^i \partial_i \nel - \alpha \nel \nabla_\mu \UU^\mu.
	\label{eq:NELEvolGeneric:app}
\end{equation}

\section{Convergence tests}  \label{app:Convergence}

We have evaluated the accuracy and the convergence properties of our code by checking how the constraint violations~\eqref{eq:CVGauss} and~\eqref{eq:CVPlasma} scale with the resolution in two test setups taken from the simulations presented in the main text. 
% We have evaluated the accuracy and the convergence properties of our code by checking how the constraint violations~\eqref{eq:GaussCartesian} scale with the resolution in two test setups taken from the simulations presented in the main text. 

% Specifically, we considered the following quantities
% \begin{align}
% CV_{\rm Gauss} &= \partial_i E^i - e \nel + \rho_{\rm(ions)}, \label{eq:CVGauss} \\
% CV_{\rm Plasma} &= \sqrt{\Gamma^2(1 - \UU_i \UU^i)} - 1, \label{eq:CVPlasma}
% \end{align}
% which, whenever nonzero, represent the violations of the Gauss law~\eqref{eq:GaussCartesian} and of the normalization condition in Eq.~\eqref{eq:PlasmaConstraintCartesian}, respectively.

% In order to asses the reliability of our code we show here the convergence 
In particular, we focus on the two most challenging nonlinear regimes: WB and blowout (although not shown, the convergence of the linear regime is excellent). Starting from the former, we repeated the simulation with $A_E = 1$ whose characteristic are described in Sec.~\ref{sec:results:nonlinear}, using a lower resolution $\Delta x = \Delta y = \Delta z = 0.4$, and increasing the grid size to $[-4, 4] \times [-4, 4] \times[-1450, 1150]$ in order to maintain $21$ grid points along the $x$ and $y$ directions. We also doubled the time step to $\Delta t = 0.2$, in order to keep the {\rm CFL} factor constant. 

\begin{figure*}[th]
    \includegraphics[width = \columnwidth]{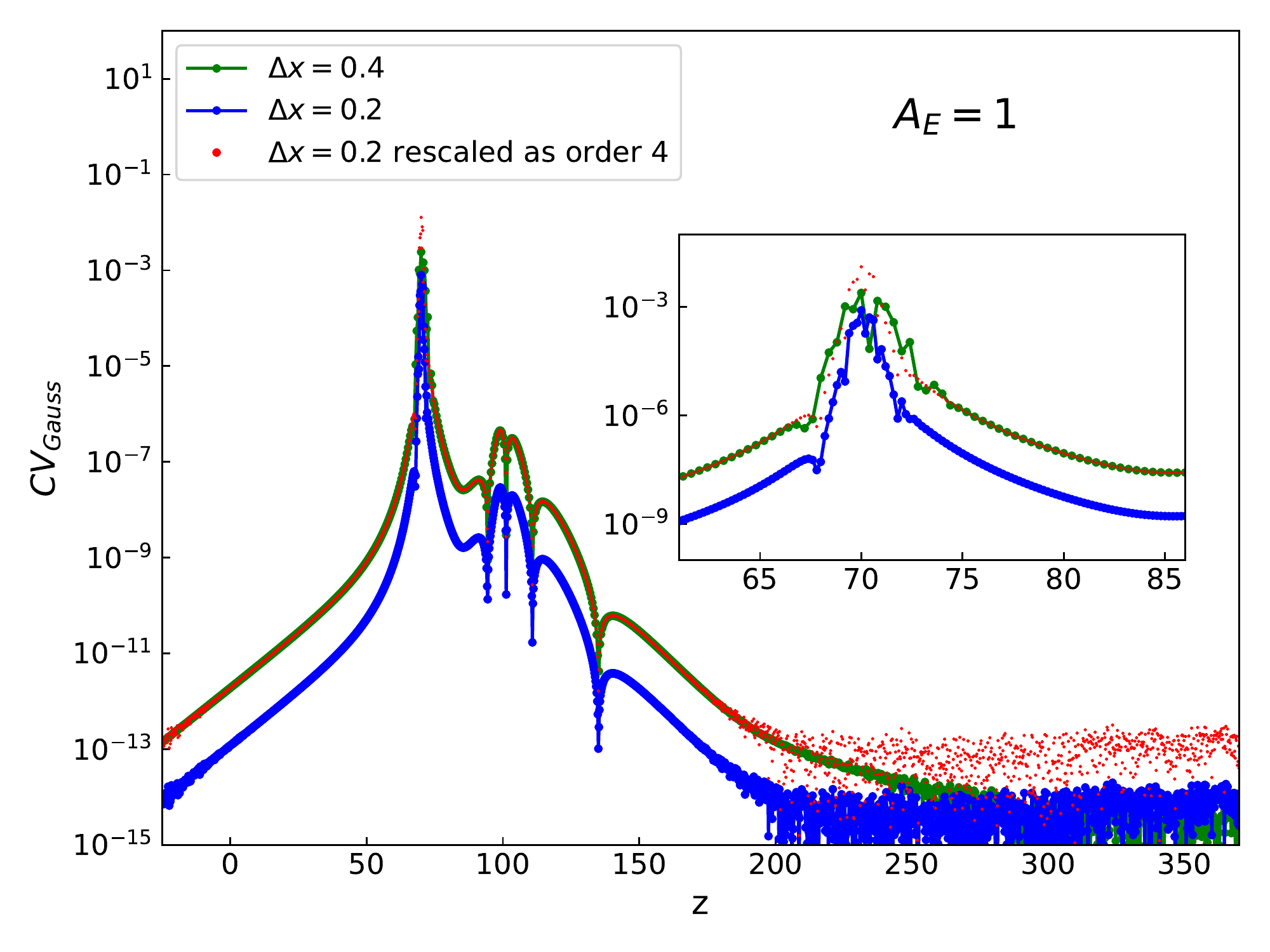}
    \includegraphics[width = \columnwidth]{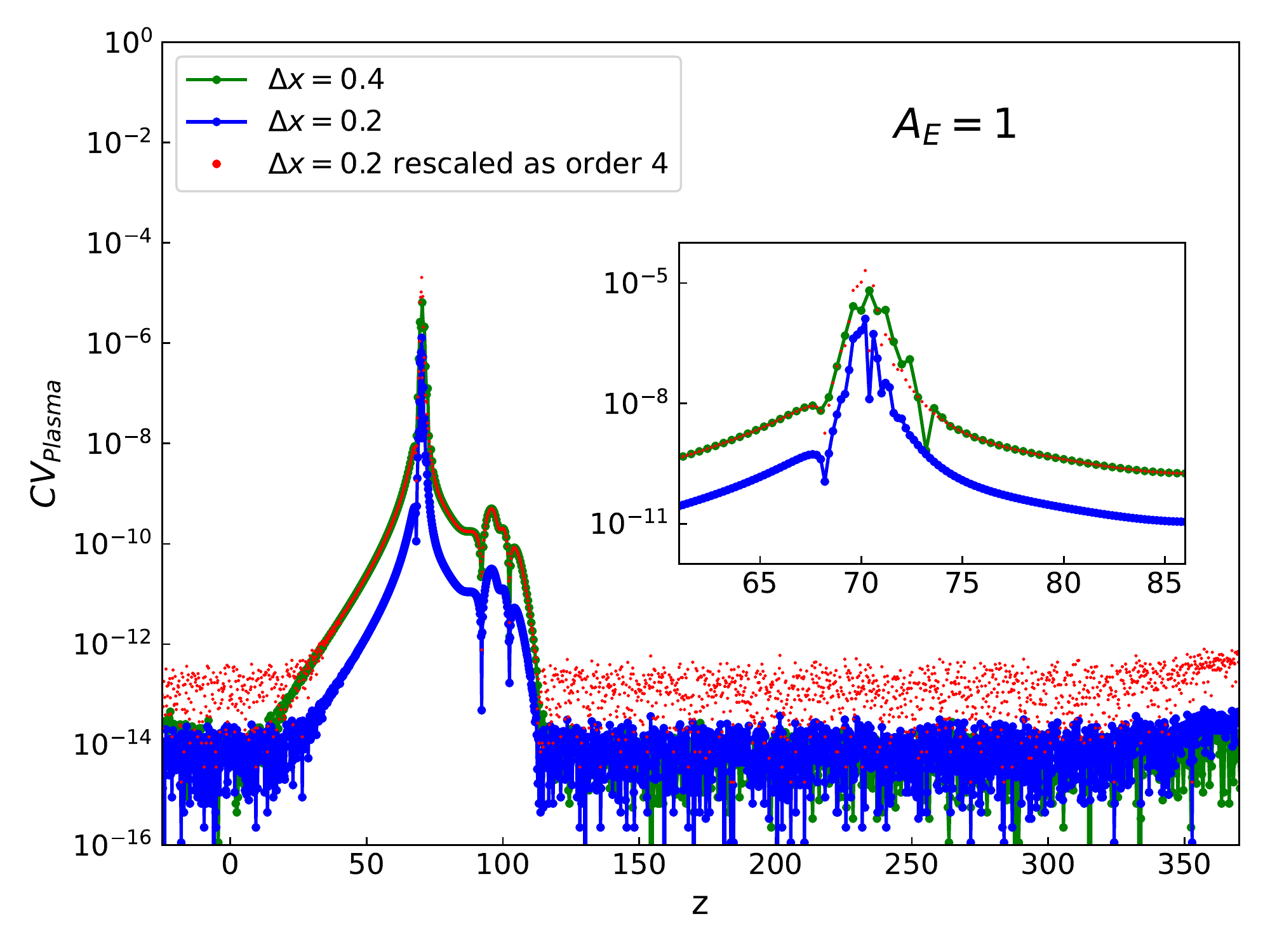}
    \caption{Scaling of the violations of the Gauss Law (left panel) and the condition $u^\mu u_\mu = -1$ along the $z$ axis, for the simulation in the nonlinear regime with $A_E = 1$. $CV_{\rm Gauss}$ and $CV_{\rm Plasma}$ are extracted at $t = 830$, when the WB phenomenon starts taking place. Overall the code converges extremely well, except in the region around the spike of $\nel$, where the constraint violation displays a peak. The insets show a magnification of the constraint violations around this region.}
    \label{fig:WaveBreakingConvergence}
\end{figure*}

Figure~\ref{fig:WaveBreakingConvergence} shows the constraint violations $CV_{\rm Gauss}$ (left panel) and $CV_{\rm Plasma}$ (right panel) along the $z$ axis at $t = 830$, slightly before WB happens (cf. lower panel of Fig.~\ref{fig:WaveBreakingAtTheThreshold}). In general, while for both the constraint violations there is a region where they are dominated by noise, in the central region they show an excellent fourth order scaling, and convergence is lost only for $65 \lesssim z \lesssim 75$, where the WB phenomenon is taking place.

\begin{figure*}[th]
    \includegraphics[width = \columnwidth]{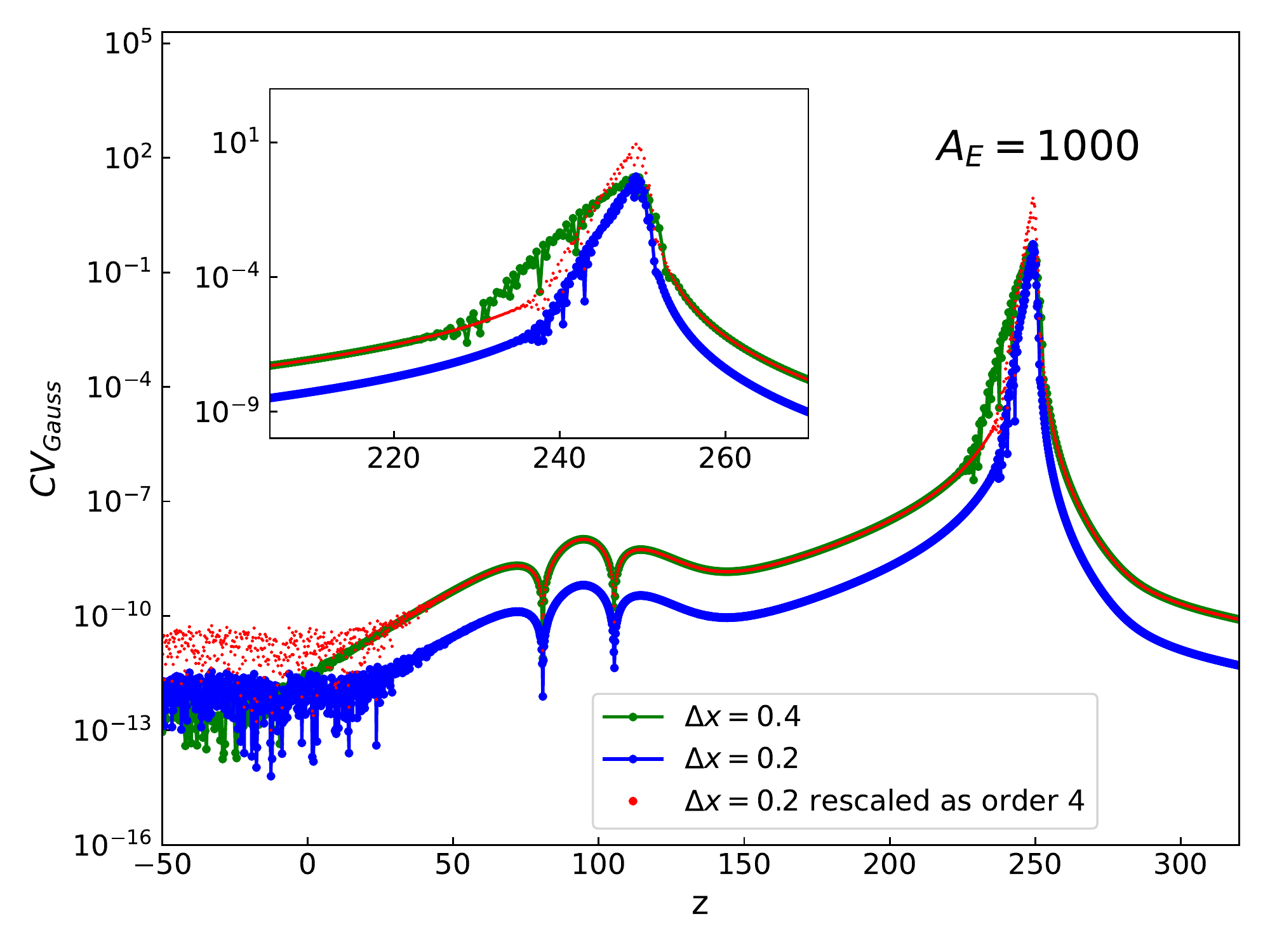}
    \includegraphics[width = \columnwidth]{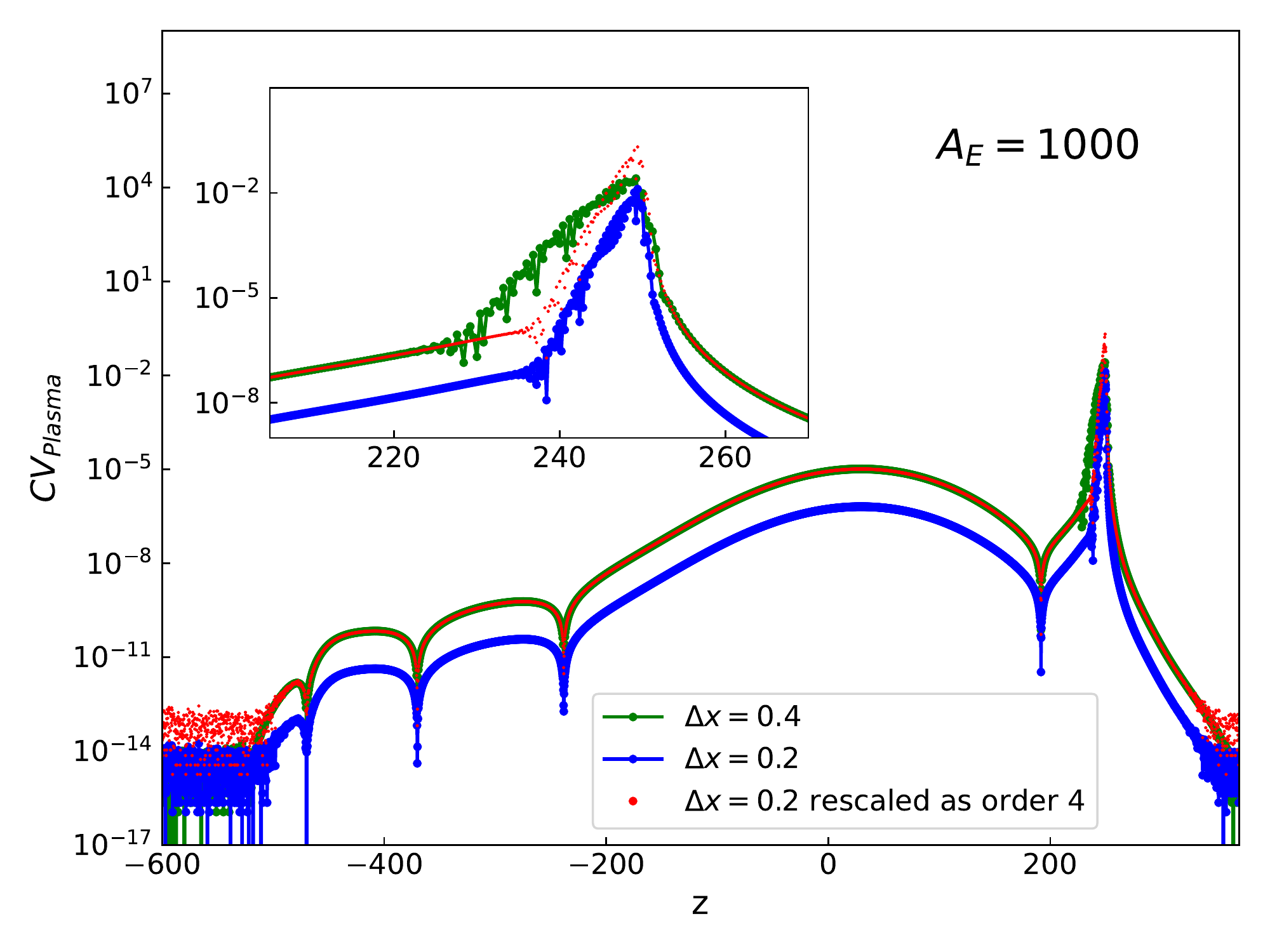}
    \caption{Convergence of $CV_{\rm Gauss}$(left panel) and $CV_{\rm Plasma}$(right panel) along the $z$ axis for the simulation in the nonlinear regime with $A_E = 1000$. The constraint violation is computed at $t = 190$, when the system is in the blowout regime. As we can see it satisfies fourth-order scaling except in the region close to the ``transported'' plasma and behind it, where the constraint violations have a peak. This can be better appreciated in the inset, that contains a magnification of the constraint violation around this region.}
    \label{fig:BlowoutRegimeCovergence}
\end{figure*}

We now move to consider the convergence in the blowout regime. We repeated the simulation with $A_E = 1000$ using grid steps $\Delta x = \Delta y = \Delta z = 0.4$ while maintaining the {\rm CFL} factor constant. As in the previous case we extended the grid to $[-4, 4] \times [-4, 4] \times[-750, 850]$ in order to have the same number of grid points along the transverse directions $x$ and $y$. 
We show the scaling of $CV_{\rm Gauss}$ and $CV_{\rm Plasma}$ on the $z$ axis at $t = 190$ in the left and right panel of Fig.~\ref{fig:BlowoutRegimeCovergence}, respectively. We can see that the code converges extremely well, except in the region just behind the peak of the plasma density (cf. lowest panel of Fig.~\ref{fig:NonLinearSnapshots}). However, we note that the extension of the region where convergence is lost decreases as the resolution increases, and that fourth-order scaling is restored in the plasma-depleted region.

Given the excellent convergence properties in the nonlinear regime, we conclude that the code is reliable and produces accurate results at the resolutions used in this work. 

\section{Homogeneity of the fields along the transverse direction} \label{app:TransverseHomogeneity}
Throughout all this work we used numerical grids whose extension along the transverse directions $x$ and $y$ is significantly smaller than in the $z$ direction. This has the advantage of reducing considerably the computational cost, and can be done by exploiting the planar geometry of the system under consideration. In this appendix, we wish to show that homogeneity of the variables along the transverse directions is preserved also at late times during the evolution, so that this grid structure is compatible with the physical properties of the system for the entire duration of the simulations. 

For this purpose we consider the simulation in the nonlinear regime with $A_E = 1000$, and we extract the profiles of $E^x$, $E^y$, $E^z$, and $\nel$ along the $x$ and $y$ axes at $z = 240$. This operation is performed at $t = 180$ when the system is already in a blowout state, and the value of the $z$ coordinate is chosen to be where plasma is concentrated at this time. 

\begin{figure*}
    \includegraphics[width = \columnwidth]{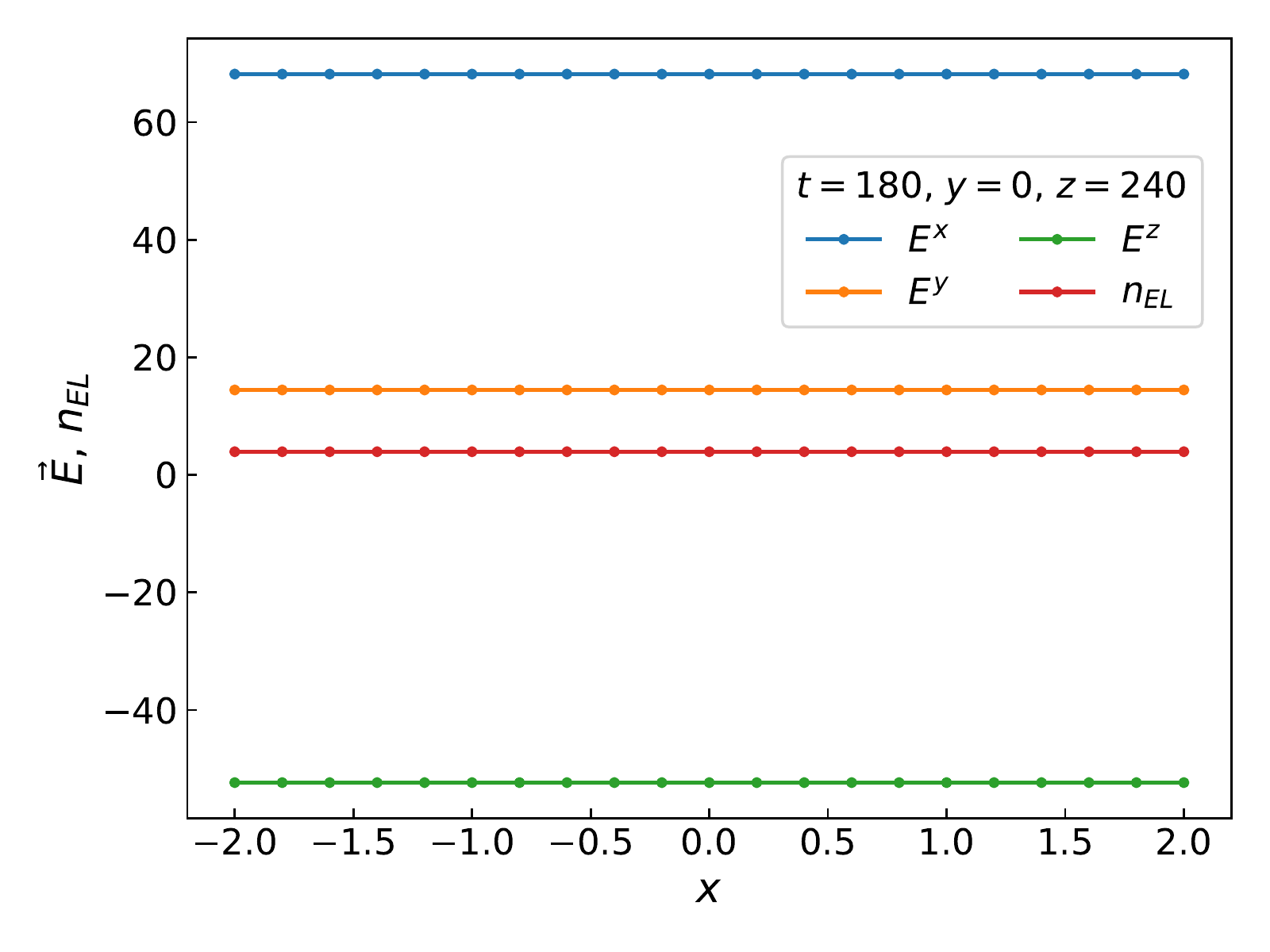} 
    \includegraphics[width = \columnwidth]{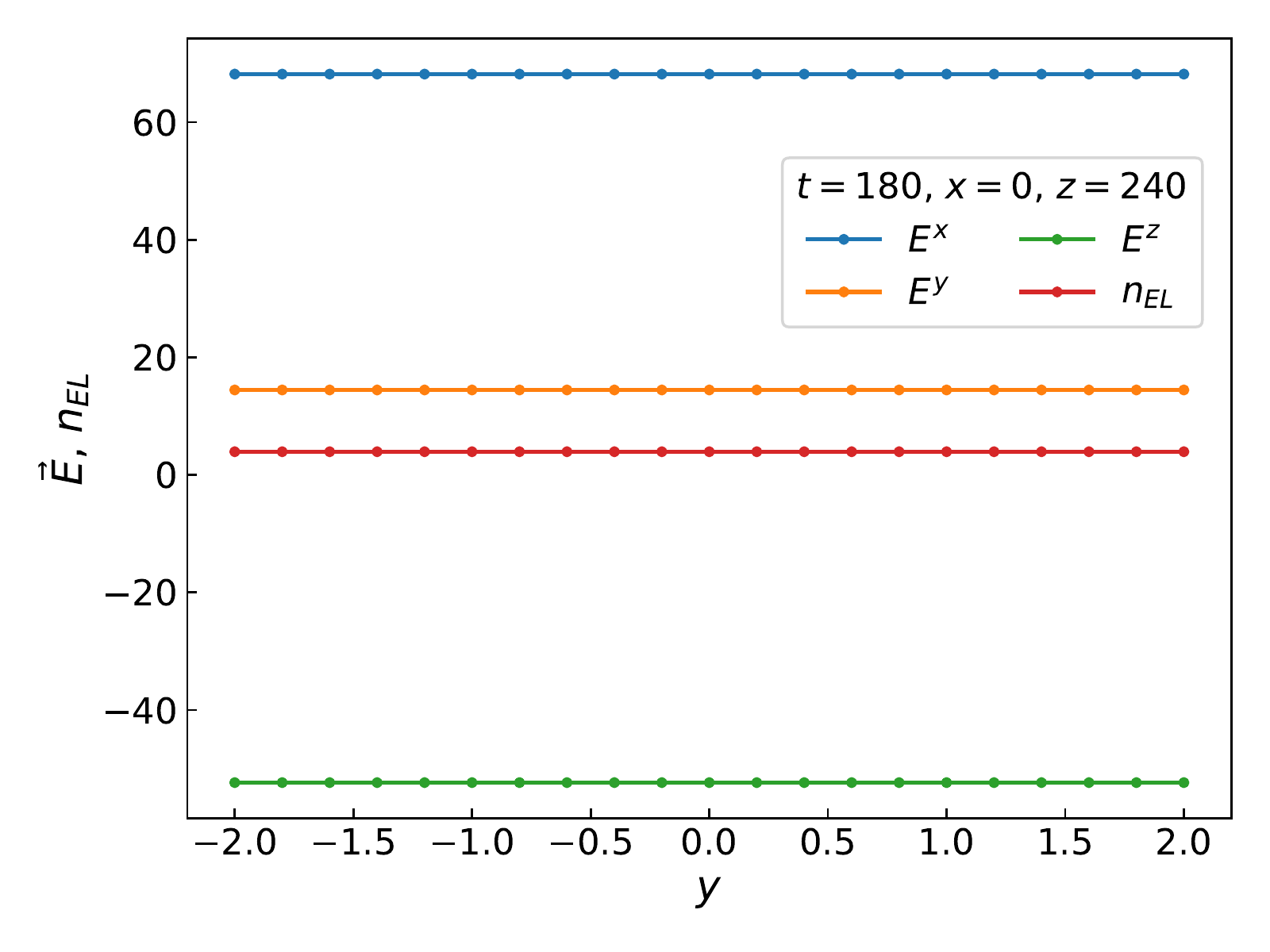}
    \caption{Profiles of $\vec E$ and $n_{EL}$ along the transverse directions $x$ (left) and $y$ (right) at $z = 240$ for the simulation with initial amplitude $A_E = 1000$. These data are extracted at $t = 180$, where the system is already in the blowout regime, and in the spatial region where the plasma density peaks. All the profiles are constant in $x$ and $y$, with values consistent between the two plots. This confirms that the homogeneity property is conserved during the $3+1$ simulations.}
    \label{fig:HomogenityTransversePlane}
\end{figure*}

We show the results in Fig.~\ref{fig:HomogenityTransversePlane}, where the left and right panels represent the profiles along the $x$ and $y$ axes, respectively. We see that all the profiles are constant along the axes, and that the values are consistent between the two plots, confirming that the system maintains homogeneity along the transverse direction.

\bibliographystyle{utphys}
\bibliography{Ref}

\end{document}